\documentclass{aastex631}
\usepackage{mathtools}

\newcommand{\vishal}[1]{{#1}}
\newcommand{\mg}{\ion{Mg}{2}}
\newcommand{\car}{\ion{C}{2}}
\newcommand{\si}{\ion{Si}{4}}

\newcommand{\bmag}{$\vert$B$\vert$}
\newcommand{\ppc}{Paper~\rm{II}}
\newcommand{\ppsi}{Paper~\rm{I}}
\received{2021 November 11}
\revised{2021 November 18}
\accepted{2021 November 23}
\submitjournal{ApJ}
\shorttitle{Origin of solar wind and coronal heating}
\shortauthors{Upendran \& Tripathi}
\begin{document}
\title{On the formation of solar wind \& switchbacks, and quiet Sun heating}

\correspondingauthor{Vishal Upendran}
\email{uvishal@iucaa.in}

\author[0000-0002-9253-6093]{Vishal Upendran}
\author[0000-0003-1689-6254]{Durgesh Tripathi}
\affiliation{Inter-University Centre for Astronomy and Astrophysics, Post Bag-4, Ganeshkhind, Pune 411007, India}
\begin{abstract}

\vishal{The solar coronal heating in quiet Sun (QS) and coronal holes (CH), including solar wind formation, are intimately tied by magnetic field dynamics}. Thus, a detailed comparative study \vishal{of these regions} is needed to understand the underlying physical processes. CHs are known to have subdued intensity and larger blueshifts in the corona. This work investigates the similarities and differences between CHs and QS in the chromosphere using the {\mg} h \& k, {\car} lines, and transition region using {\si} line, for regions with identical absolute magnetic flux density ({\bmag}). We find CHs to have subdued intensity in all the lines, with the difference increasing with line formation height and {\bmag}. The chromospheric lines show excess upflows and downflows in CH, while {\si} shows excess upflows (downflows) in CHs (QS), where the flows increase with {\bmag}. We further demonstrate that the upflows (downflows) in {\si} are correlated with both upflows and downflows (only downflows) in the chromospheric lines. CHs (QS) show larger Si IV upflows (downflows) for similar flows in the chromosphere, suggesting a common origin to these flows. These observations may be explained due to impulsive heating via interchange (closed-loop) reconnection in CHs (QS), resulting in bidirectional flows at different heights, due to differences in magnetic field topologies. Finally, the kinked field lines from interchange reconnection may be carried away as magnetic field rotations and observed as switchbacks. Thus, our results suggest a unified picture of solar wind emergence, coronal heating, and near-Sun switchback formation. 

\end{abstract}

\section{Introduction} \label{sec:intro}
The temperature of the solar atmosphere varies from $\approx5500$~K at the photosphere to $\geq10^6$~K in the corona. The intervening chromosphere variably lies at around $\approx10^4$~K, with a steep rise occurring in the transition region (TR). The solar atmosphere further extends outward from the corona, filling the heliosphere with the streaming solar wind. All these layers of the solar atmosphere interact through a continuous exchange of mass and energy, and are tightly coupled by the highly dynamic magnetic field. Various processes have been proposed, and in some cases, observed to be the drivers of this mass and energy transport. The two primary processes thought to be responsible are dissipation of MagnetoHydroDynamic (MHD) waves \citep[see e.g.][]{alfven_1947_waveheating}, and that of currents produced due to the braiding of magnetic field lines \citep[see, e.g.][and also \cite{klimchuk_2006_coronalheating, parnell_2012coronalheatingreview} for comprehensive reviews]{parker_1972_Bfieldbraiding, tzihong_1987_reconnection, Parker_1988_nanoflares}.

In the solar corona, we usually find three morphologically different regions, {\it viz.} Coronal Holes (CHs), Active Regions (ARs) and Quiet Sun (QS). The CHs are features seen as dark structures in Extreme UltraViolet (EUV) and X-ray images of the solar corona, while ARs are seen as localized bright structures. In addition to these dark and bright structures, the omnipresent background over which the CHs and ARs are observed is called the QS. Note that while CHs are clearly distinguishable from QS in the EUV and X-ray images, these two regions appear extremely similar at lower heights {\it viz} the chromosphere and photosphere \citep[see e.g.,][]{Stucki_FUVLinesSumer,Stucki_UVLinesSumer_Chs,PradeepKashyap2018,TriNS_2021,Upendran_C2}. However, note that the \ion{He}{1} 10830~{\AA} (an absorption line) shows excess intensity in CHs \citep{harvery_heI10830,Kahler_heI10830}, while the \ion{He}{1} 584~{\AA} (an emission line) shows lower intensity \citep{Jordan_heI584}, and excess blueshift, line width in the CHs, over QS \citep{Peter_1999_HeICHQS}. However, these differences may be attributed to the sensitivity of these lines to coronal radiation, reflecting conditions in the corona. Furthermore, at 17~GHz in microwave, CHs are found to be brighter than QS \citep{gopalswamy1999_CHmicrowave}, while this difference is not observed in radio wavelengths at 1.2~mm \citep{brajvsa2018_CHradio}. Thus, a gross differentiation of a given region in QS or CH is markedly seen predominantly in the coronal observations.

Observations by \cite{Krieger1973} indicated that the high velocity streams of the solar wind may be traced back to the CHs on the Sun. The ``slow'' wind, on the other hand, has been variously traced to the edges of ARs~\citep[][]{brooks_2015_slowwindsources} and equatorial CHs~\citep[see e.g.][]{Bale_2019_switchbacke}. Similar results have been obtained by various authors \citep[see e.g.][]{wang_1990_WSAENLIL, Schwenn_2006_SWSources, JanTM_2008} and more recently through the application of Deep learning based localization by \cite{Upendran_solarwind}. These results demonstrate that CHs are potential source regions of the solar wind and thus provide an opportunity to investigate the physical processes involved in the formation of solar wind. Since CHs and QS hardly differ at the lower atmospheric heights, studying them in tandem may allow us to explore the possibility of a unified explanation of heating in the QS and CHs, including the formation of solar wind \cite[][]{TriNS_2021}.

\cite{Hassler810} investigated differences between CH and QS using the \ion{Si}{2} and \ion{Ne}{8} lines, and found a relation between the network regions and blueshifts of \ion{Ne}{8}, with more blueshifts in CHs. Comprehensive studies of CHs and QS were undertaken by \citet[][]{Stucki_UVLinesSumer_Chs,Stucki_FUVLinesSumer} using spectral lines sensitive to a range of temperatures from $\approx8\times10^3$~K to $\approx1.4\times10^6$~K. While CHs showed a clear deficit in intensity, excess blueshift and excess line width with respect to QS for spectral lines forming at a temperature higher than $\approx4\times10^5$~K, at chromospheric temperatures the differences were negligible, and within the measurement error. Similarly, \cite{xia_chsumer} studied the relationship between Doppler shifts of {\car}, \ion{H}{1} Ly$\beta$, and \ion{O}{6} in CHs, and found a direct relation between the Doppler shifts of \ion{O}{6} with that of {\car} and \ion{H}{1} Ly$\beta$. These correlated shifts led \cite{xia_chsumer} to conclude that these shifts are signatures of solar wind in the chromosphere. However, note that the associated uncertainties in the velocity scatter obtained by \cite{xia_chsumer} were large. Moreover, while the average chromospheric velocities in bins of the \ion{O}{6} velocities were studied, the systematic associations between red and blue shifted pixels, separately, for these lines, was not performed by \cite{xia_chsumer}. 

The correspondence between network region and outflows in the CHs, using \ion{Ne}{8} line, demonstrated by \cite{Hassler810} was further investigated by \cite{tu2005solar}, by mapping the formation heights of \ion{Si}{2}, \ion{C}{4} and \ion{Ne}{8} in a CH. On further detailed investigation, \cite{tu2005solar} showed a clear relation between the \ion{Ne}{8} blueshifts and the underlying magnetic field configuration, obtained using the potential field extrapolation of photospheric magnetic field. Thus, \cite{tu2005solar} suggested a modulation of the solar wind velocities due to the underlying magnetic field configuration.

More recently, \cite{PradeepKashyap2018} investigated the intensity differences between CH and QS in the {\mg}~k line, observed by the Interface Region Imaging Spectrometer \citep[IRIS,][]{iris}. They find a clear deficit of intensity in CHs over QS for regions with similar absolute photospheric magnetic flux density ({\bmag}), and with larger difference for larger {\bmag}. Similar analysis for the intensity, velocity and non thermal widths for \ion{Si}{4} was performed by \citet{TriNS_2021} (henceforth referred to as {\ppsi}). Similar to the results of \cite{PradeepKashyap2018}, intensity deficit in CHs over QS for regions with similar {\bmag} was observed. Moreover, CH (QS) was more blueshifted (redshifted) over QS (CH) for identical {\bmag}. However, no significant difference was observed in the non-thermal width between CH and QS. The excess CH blueshifts were interpreted to be signatures of nascent solar wind at {\si} formation heights in {\ppsi}. Thus, while a clear signal of solar wind was reported in the hotter \ion{Ne}{8} line by \cite{tu2005solar}, the signatures are already present in the upper TR line \ion{Si}{4}, if the underlying photospheric magnetic flux density distribution is taken into account. Furthermore, since the regions with identical {\bmag} were compared, the deficit in intensity in CHs over QS would mean energy to be either used to accelerate the solar wind, or heat up the corona. Thus, a unified picture of solar wind formation \& coronal heating was presented in {\ppsi}.

In a companion paper \citep[][hereafter referred to as {\ppc}]{Upendran_C2}, we perform investigation, similar to {\ppsi}, on the {\car} $1334$~{\AA} line. We find intensity deficit and excess blueshifts in CHs over QS for regions with identical {\bmag}, similar to the results from {\ppsi} in {\si} line. However, we also find excess redshifts in CHs over QS for regions with identical {\bmag}, which is the opposite of what is observed in the {\si} line. Finally, we find the total line width to be larger in CH over QS for regions with similar {\bmag}, while the line skew \& kurtosis were found to be similar in both CH and QS, suggesting that similar physical processes are at work in the two regions, as was also concluded in {\ppsi}.

All these investigations, taken together, raise several questions: Where does the solar wind actually originate? Can a single height be even ascribed to it? At what height does the differentiation between CH and QS start? Does the solar wind emergence have any relation to the origin of a million degree Kelvin corona? These questions are critical to diagnose the formation signatures of the solar wind. Furthermore, attempting to answer these questions will also narrow down on the viability and contribution of various heating mechanisms in the solar atmosphere.

Keeping the above questions in mind, we first study the {\mg} h \& k line dynamics in CH and QS. We then go ahead and explore the correlations between the {\mg}, {\car} and {\si} lines in CH and QS, with the intention to explain the observations. The remainder of the paper is structured as follows: In \S \ref{sec:data}, we describe our observations, with feature extraction in \S\ref{sec:featureextract}. In \S\ref{sec:mg}, we present results of the {\mg} lines, while we recapitulate results for the {\car} and {\si} lines from {\ppc} and {\ppsi} in \S\ref{sec:c2feature} and \S\ref{sec:si4feature}. In \S\ref{sec:combined}, we summarize and discuss our results. Finally, we provide an interpretation on context of origin of solar wind, switchbacks \citep{Bale_2019_switchbacke} and QS coronal heating \S\ref{interpretation}.

\section{Data} \label{sec:data}
\begin{deluxetable*}{|c|c|c|c|c|}
  \tablecaption{Details of the IRIS rasters used in this study. Note that the dataset used is the same analyzed in {\ppc}, and the average $\mu$ is mentioned for each Field of View. \label{tab:datadetails}}
  \tablewidth{0pt}
  \tablehead{\colhead{Dataset name} & \colhead{Time range} & \colhead{(Xcen,Ycen)} & \colhead{Raster FOV} & \colhead{$\mu$}}
   \startdata
  DS1 & 2014-07-24 11:10:28 -- 14:40:53 & (128{\arcsec},-180{\arcsec}) & (141{\arcsec},174{\arcsec}) & 0.97 \\
  DS2 & 2014-07-26 00:10:28 -- 03:40:53 & (469{\arcsec},-167{\arcsec}) & (141{\arcsec},174{\arcsec}) & 0.85\\
  DS3 & 2014-08-02 23:55:28 -- 03:25:53 +1d & (332{\arcsec},-152{\arcsec}) & (141{\arcsec},174{\arcsec}) & 0.92\\
  DS4 & 2015-04-26 11:39:31 -- 15:09:56 & (-288{\arcsec},45{\arcsec}) & (141{\arcsec},174{\arcsec}) & 0.95\\
  DS5 & 2015-10-14 11:07:33 -- 14:37:58 & (215{\arcsec},-165{\arcsec}) & (141{\arcsec},174{\arcsec}) & 0.97\\
  \enddata
\end{deluxetable*}
In this study, we use the observations recorded by IRIS, the Atmospheric Imaging Assembly~\citep[AIA;][]{BoeEL_2012} and the Helioseismic and Magnetic Imager~\citep[HMI;][]{HMI}. AIA and HMI are both on board the Solar Dynamics Observatory~\citep[SDO;][]{SDO}. IRIS observes the Sun in two modes, {\it viz.} slit-jaw imaging (SJI) and spectroscopy. In the SJI mode, IRIS provides photometric context images in NUV and FUV with a pixel size of $\approx0.16${\arcsec}, and an approximate cadence of $\approx63$~s. We consider the SJI data centered around 1330~{\AA} and 2796~{\AA} for co-alignment purposes. The spectroscopy is performed in three windows, one in near ultraviolet (NUV) from 2782.7 to 2851.1~{\AA}, and two in the far ultraviolet (FUV), from 1332 to 1358~{\AA} (FUV 1), and from 1389 to 1407~{\AA} (FUV 2). These have a pixel size of $\approx0.16${\arcsec} along the slit, and sample at $\approx.33${\arcsec} across the field of view (FOV). The spectral pixel size in these rasters is $\approx25.9$~m{\AA}. Time cadence between successive slit positions is $\approx30$~s. For further details on IRIS, see \cite{iris}.

AIA observes the Sun's atmosphere in UV and EUV using eight different passbands sensitive to plasma at different temperatures \citep{ODwyer, BoeEL_2012}. Here, we use the 193~{\AA} images to distinguish between CHs and QS, and the 1600~{\AA} images to co-align the IRIS, AIA and HMI observations, as described in {\ppc}. We obtain the information on the photospheric absolute magnetic flux density (i.e. {\bmag}) from the line-of-sight (LOS) magnetograms obtained with HMI. The AIA images are taken with a pixel size of $\approx$0.6{\arcsec}, with the EUV images a time cadence of $\approx12$~s, while the Ultraviolet images are taken at $\approx24$~s cadence. HMI obtains the $B_{LOS}$ magnetograms at $\approx45$~s cadence with a pixel size of 0.5{\arcsec}.

For our study, we have analyzed five different sets of observations recorded by IRIS in spectroscopic mode. The main criteria used to select these observations are that the raster must include CH and QS within the same FOV, and that they must be taken within latitude and longitude of $\pm$60$^\circ$. The IRIS observation details are given in Table.~\ref{tab:datadetails}. Note that the same data-set have been analyzed in {\ppc} to the study the properties of {\car} lines in CHs and QS. Out of these, three of the observations {\it viz.} DS1, DS2 and DS5 were also studied in {\ppsi} to characterize the similarities and differences in QS and CHs in TR using {\si} line. We use corresponding coordinated AIA data cubes, with cutouts from the full disk data used from HMI. 

The {\mg} h \& k lines form near 2803.53{~\AA} and 2796.35{~\AA}, respectively, while the {\car} and {\si} lines form near 1334.53{~\AA} and 1393.755~{\AA}, respectively. The {\mg} and {\car} lines form in an optically thick chromosphere under non local thermodynamic equilibrium conditions \citep[see, for e.g.][]{leenaarts_iris2,Rathore_CII_paper2}. Thus, these lines show extremely complex features and have non-trivial associations with local plasma properties. They have been explored in detail in {\ppc}, and \cite{Rathore_CII_paper2,leenaarts_iris2}. For all practical purposes, the {\si} line, forming in QS TR, can be considered to be formed in optically thin conditions \citep{TriNID_2020,gontikakis_si4} and its properties in QS and CH are studied in detail in {\ppsi}.

\subsection{Feature extraction} \label{sec:featureextract}
The {\mg} lines offer crucial information on the plasma conditions in the formation region, encoded into the line intensities and Doppler shifts of the line core (k3 \& h3) and the peaks (k2v, k2r, h2v \& h2r). For a detailed analysis and discussion of these lines, see \cite{leenarts_iris1,leenaarts_iris2,Tiago_iris3}. 

We first extract the positions and intensities of these different spectral line features. For this purpose, we develop a peak finding algorithm based on \cite{leenaarts_iris2,Tiago_iris3} that locates the zero-crossing of $\mathrm{dI}/\mathrm{d\lambda}$ within a window of $\pm~40$~km~s$^{-1}$ from the reference wavelength \citep[taken to be 2796.350~{\AA} and 2803.529~{\AA} for the k and h lines respectively, see][]{Tiago_iris3}.

The line core is identified to be the location with minimum intensity at the zero crossing. If the procedure is unable to locate such a minimum, e.g., in case of single-peaked or noisy profiles, we assign a default velocity of 5~km~s$^{-1}$ following \cite{leenaarts_iris2}, since the remaining procedure rests on the identification of line core. Note, however, that the {\mg} spectral profiles in this study i.e., for QS and CHs, are predominantly double-peaked, as also noted by \cite{leenaarts_iris2}.

The two peaks closest to the line core on either side are considered as the k2 (h2) peaks. Since the line core and peaks form at local extrema of the line profiles (as a function of wavelength), they may be approximated to be a parabola close to the peak value. Thus, we may fit a parabola near the maximum/minimum, and obtain a better estimate of the real extremum. This is called sub-pixel centroiding~\citep[similar to][]{teague_2018_subpixelcentroid}. Thus, the velocities and intensities for the core and peaks are then determined by fitting a parabola to the points near the feature extremum. Profiles that contain missing values of $-200$ are discarded. This procedure provides us with the intensities and Doppler shifts of the peaks \& core of {\mg} h \& k lines. The line peak Doppler shifts are determined by taking the signed average of shifts of the blue and red peak \citep{leenaarts_iris2}. 

\begin{figure}[ht!]
\centering
  \includegraphics[width=0.9\textwidth]{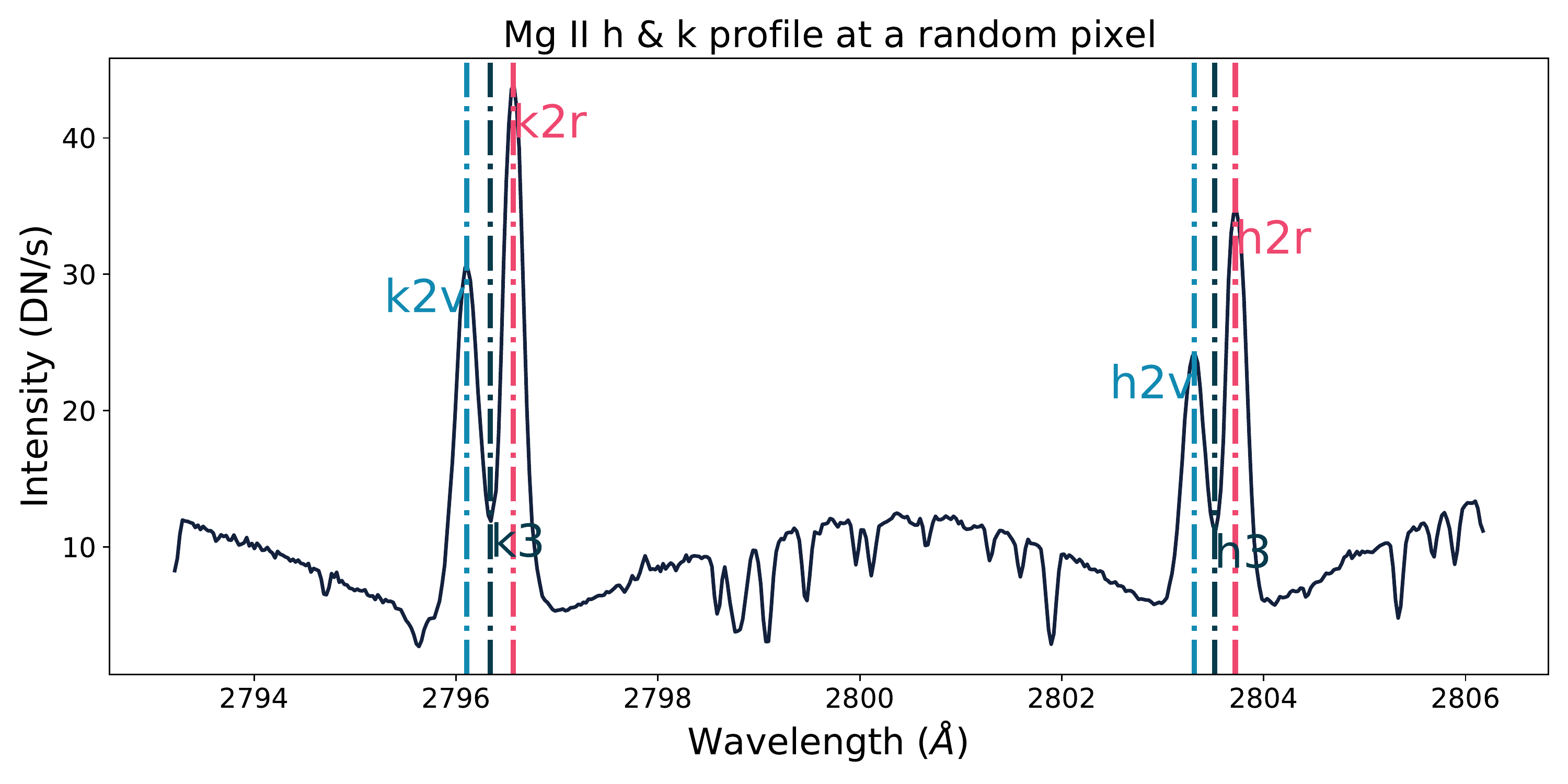}
  \caption{{\mg} spectrum at a random QS location of DS4. The {\mg} line features, along with their locations are labeled as k2v or h2v (blue), k3 or h3 (black) and k2r or h2r (red).}
  \label{fig:rasterfov}
\end{figure}

Fig.~\ref{fig:rasterfov} displays a spectrum obtained at a random pixel in DS4 centered at the two {\mg} lines. The two lines and their associated features are labeled. The core \& peaks have been identified using the algorithm presented above. The black vertical line denotes the line core. The red (blue) vertical line corresponds to red (blue) peak of the line. This convention is followed for both the h and the k lines.

Since this paper aims to investigate the dynamics of CHs and QS at different heights in the solar atmosphere, we shall first present the analysis and results obtained from one dataset (DS4) for {\mg} lines in \S\ref{mg2_1}. In the end, as has been done in {\ppsi} and {\ppc}, we average the results obtained for all five data sets in \S\ref{mg2_all}. Since the same dataset is studied in \ppc, we import the final results for the {\car} lines from \ppc. For the {\si} line, we present results obtained from the extended dataset based on the analysis performed in {\ppsi}.

\section{Results from the analysis of the \texorpdfstring{\mg}. lines}
\label{sec:mg}
\subsection{\texorpdfstring{\mg}:: Single dataset}\label{mg2_1}

\begin{figure*}[!htpb]
\centering
  \includegraphics[width=0.9\textwidth]{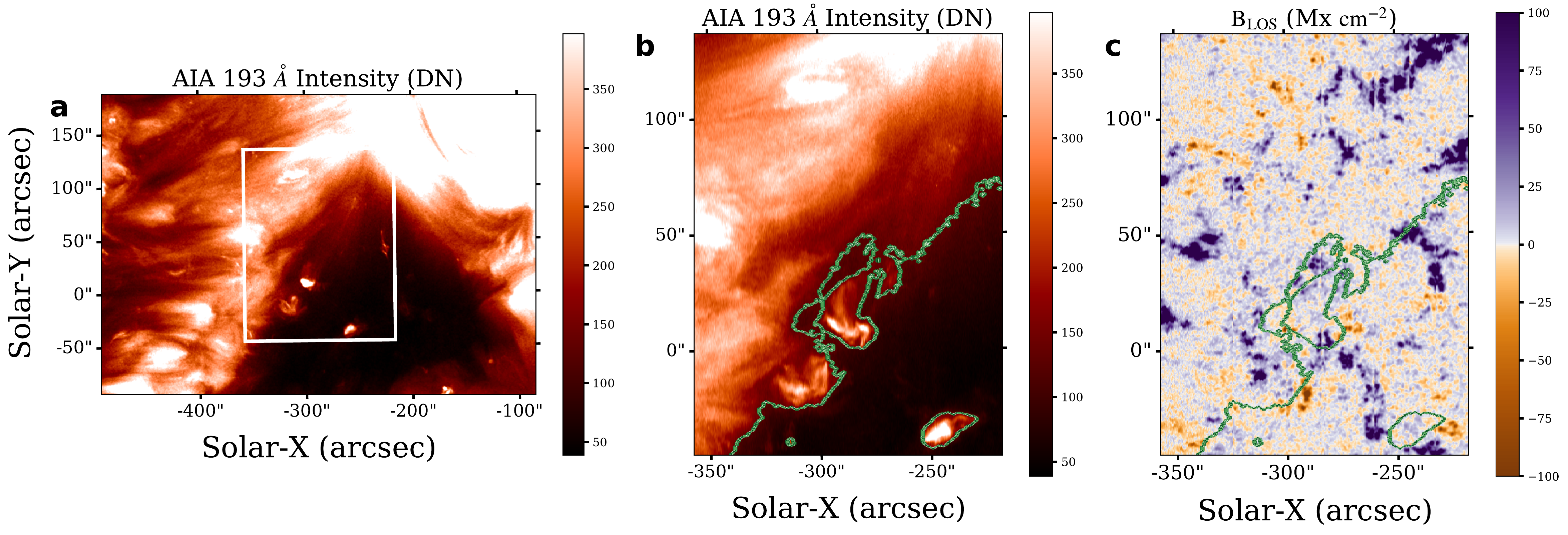}
  \caption{AIA 193~{\AA} context image (Panel~\textbf{a}). The over-plotted white box corresponds to the IRIS raster FOV. The pseudo-rasters obtained from 193~{\AA} images and HMI LOS magnetograms corresponding to the IRIS raster for DS4 are shown in panels~\textbf{b} \& \textbf{c}, respectively. The green contours in panels \textbf{b} and \textbf{c} demarcate the CH and QS, obtained from the segmentation algorithm.}
  \label{fig:segmentation_profiles}
\end{figure*}

Fig~\ref{fig:segmentation_profiles}.\textbf{a} displays a portion of the solar disk obtained from AIA~193~{\AA} full disk image. The over-plotted white box represents the IRIS raster FOV. Panels~\textbf{b} and \textbf{c} display the pseudo-rasters of AIA 193{~\AA} and HMI LOS magnetogram. Similar to {\ppc}, we apply the segmentation algorithm from \cite{Upendran_solarwind} to the AIA 193~{\AA} pseudo-rasters to obtain a demarcation of CHs from QS. In Fig.~\ref{fig:segmentation_profiles}.\textbf{b} \& \textbf{c}, the green contours demarcate CH from QS. We clearly see that the HMI pseudo-raster does not show any visual difference between CHs and QS, similar to the results obtained by \cite{TriNS_2021, PradeepKashyap2018}.

We now investigate the dependence of the following features on {\bmag} through scatter plots: 1) core \& peak intensities of the two lines, 2) intensity ratios of the two peaks, 3) line core velocities and 4) average peak velocities. Note that we consider 10~G as the noise floor of {\bmag} \citep{yeo_10gauss_bfielderror,couvidat2016observables}.  

Following the procedure outlined in {\ppsi} and {\ppc}, to improve the signal to noise ratio (SNR) and statistics, we consider the derived quantities in the bins of {\bmag}, and report the average values in these bins. We use a constant {\bmag} bin size of 0.1 in log-space to account for the fewer pixels at high {\bmag}. Note that the LOS {\bmag} and Doppler shifts are converted to radial field and flows by dividing with $\mu$ \citep[the heliocentric coordinate, see][]{2006_Thompson_muangle} of the respective pixel. Furthermore, the errors reported in all the plots are the standard errors on the mean. The standard error is defined as $\sigma/\sqrt{N}$, where $\sigma$ is the standard deviation for the samples present in the bin, and $N$ is the number of samples. Note that while we are interested in and report the variation of mean value in each bin, we present the distribution of samples in each bin with $1$ and $90$ percentile bounds in the Appendix. \ref{sec:appdistr}.
\subsubsection{Intensities}\label{sec:mg2intensity}
\begin{figure*}[htp!]
  \centering
  \includegraphics[width=0.8\linewidth]{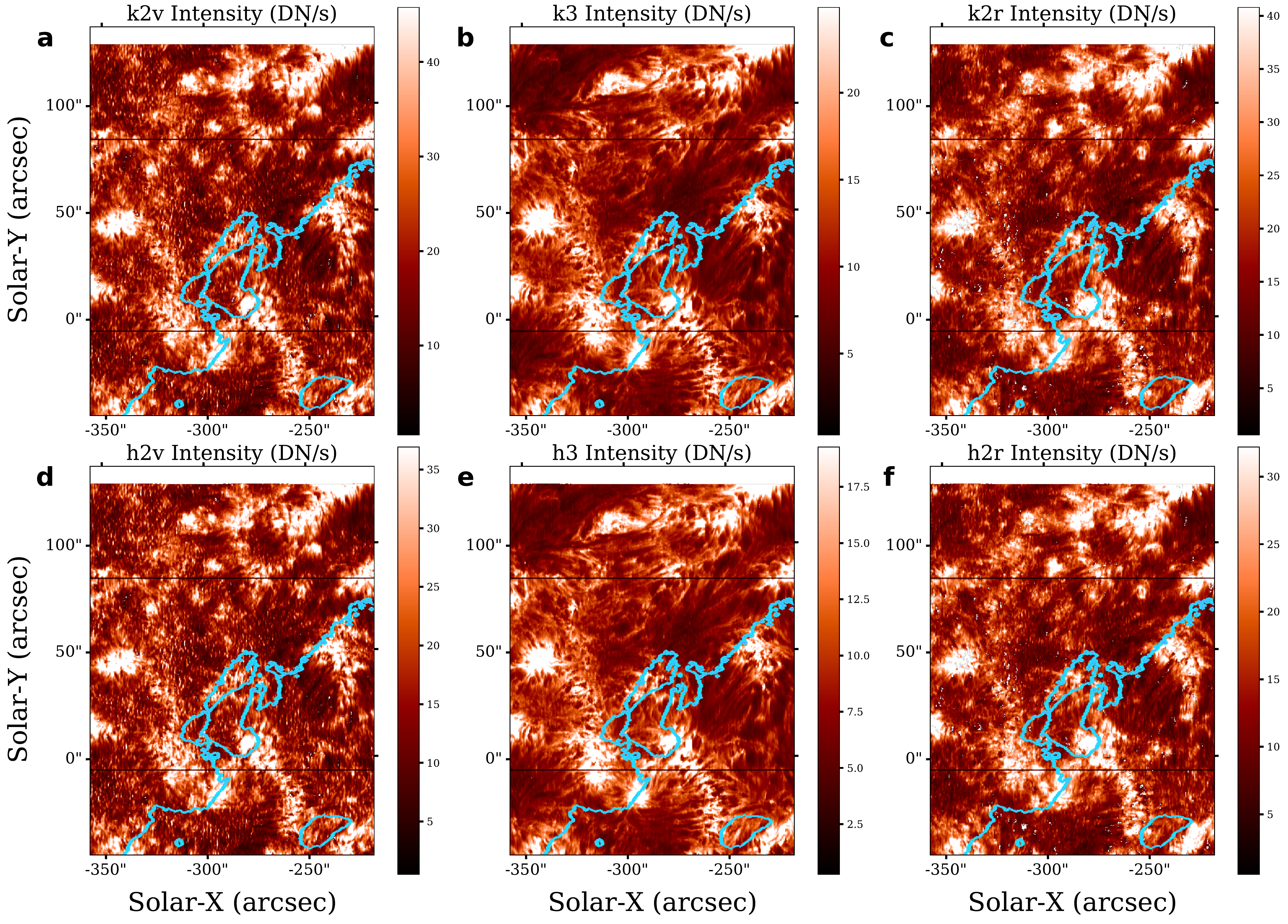}
  \caption{Intensity maps obtained in {\mg} k (top row) and h (bottom row) line features from DS4. The blue contours represent CH-QS boundary, as shown in panel Fig.~\ref{fig:segmentation_profiles}.\textbf{b}.} \label{fig:mg2intens_map}
\end{figure*}
\begin{figure*}[htp!]
  \centering
  \includegraphics[width=\linewidth]{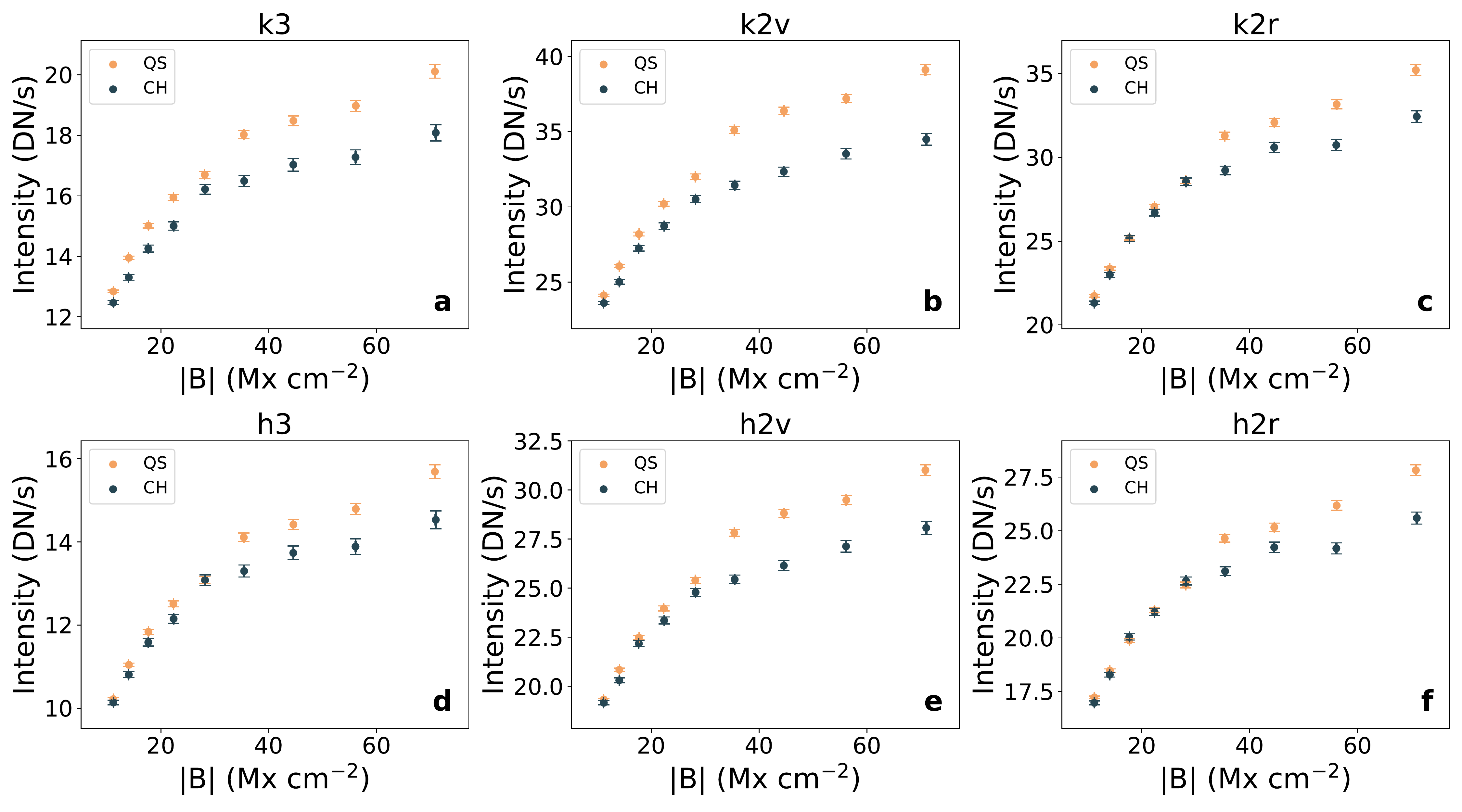}
  \caption{Variation of intensities in {\mg} k (top row) and h (bottom row) line features with {\bmag}. The orange color indicates QS, and black indicates CH. Note that the standard errors in {\bmag} have also been plotted in this figure, and all subsequent figures, but they are too small to be seen.}
  \label{fig:mg2intensity}
\end{figure*}
\begin{figure*}[!ht]
  \centering
  \includegraphics[width=0.8\textwidth]{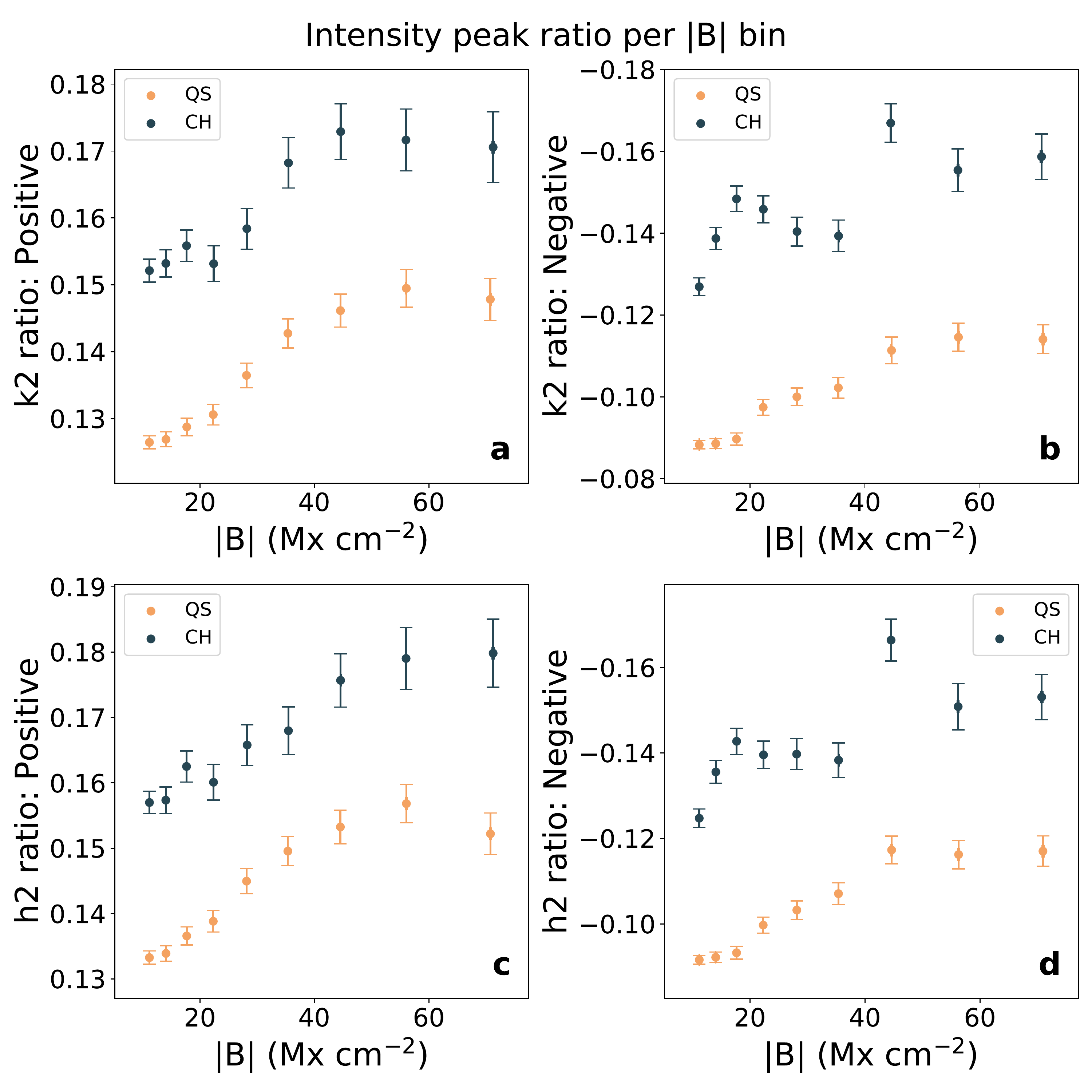}
  \caption{The peak ratios as a function of {\bmag} for the k (top row) and h (bottom row) lines. The positive ratio (downflowing plasma) are depicted in panels \textbf{a} \& \textbf{c}, while the negative ratio (upflowing plasma) are depicted in panels \textbf{b} \& \textbf{d}. Note the absolute values of the ratio increase along the y-axis for all the plots.}
  \label{fig:mg2peakratio}
\end{figure*}

First, we consider the intensities obtained from the two {\mg} lines. We have six intensity measurements in total; four from the peaks, and two from the cores of h \& k lines. In Fig.~\ref{fig:mg2intens_map}, we display the intensity maps obtained in these features for DS4. The over-plotted blue contours demarcate the QS and CH. We clearly see no visible difference between CH and QS in any of the features of {\mg} line. However, a clear relation is seen with the photospheric magnetic flux density in Fig.~\ref{fig:segmentation_profiles}.\textbf{c}, inline with the results of \cite{PradeepKashyap2018} for {\mg}~k line.

In Fig.~\ref{fig:mg2intensity}, we plot the intensities of different {\mg} h\& k features in bins of {\bmag}. In the plots, black (orange) data points represent CH (QS), with the k (h) line features in the top (bottom) row. We clearly see that the intensity increases with {\bmag} for both CH and QS for all the line features. Furthermore, the QS shows excess intensity over CH for {\bmag} $\ge$30~Gauss. However, there is a mild difference in the intensities already at 10~G for the k line. We further note that the difference in intensities between QS and CH increases with increasing {\bmag}, with an apparent saturation at higher {\bmag}. These results are in agreement with those reported by \cite{PradeepKashyap2018}.

Another key inference from Fig.~\ref{fig:mg2intensity} is the larger intensities of the blue peaks (k2v and h2v) over the red peaks (k2r and h2r; see panels \textbf{b}, \textbf{c}, \textbf{e} and \textbf{f}). Note that the peak ratio (I$_v$-I$_r$)/(I$_v$+I$_r$) is a proxy for the average chromospheric velocity, as has been suggested by \cite{leenaarts_iris2}. Positive peak ratio corresponds to down-flowing plasma in the atmosphere, while a negative ratio corresponds to up-flowing plasma. A preferentially larger blueward or redward peak arises due to increased absorption on the side of the smaller peak~\citep[see][for details]{leenaarts_iris2}. The enhanced intensities in the blue peaks over red peaks suggests that the chromosphere is largely redshifted, resulting in increased redward absorption at the height corresponding to {\mg} formation. Note that unless stated otherwise, redshift means plasma moving towards the Sun and blueshift means plasma moving away from the Sun.

In the following, we consider pixels with only positive and negative ratios separately and the variation of the ratio with {\bmag}. This would amount to considering only pixels which have downflows (or upflows), as a function of {\bmag}. Fig.~\ref{fig:mg2peakratio} plots positive (panel \textbf{a} \& \textbf{c}) and negative ratios (panels \textbf{b} \& \textbf{d}) for k2 and h2 line features. From the plots, we find that the peak ratios vary between 0.1 and 0.2, which is in a sufficiently linear regime of the scatter between peak ratio and average $v_z$ \citep[as may be seen in Fig. 8.e and f of][]{leenaarts_iris2}. Thus, we may consider the peak ratio as a proxy for the average chromospheric velocities in CH and QS. The plots further show that the peak ratio becomes increasingly positive or negative with rising {\bmag} till 50 Gauss and saturates thereafter. Also note that the positive as well as negative peak ratios are larger in CHs than in QS for identical {\bmag}. This intriguing finding is indicative of larger downflows as well as upflows in CH over QS for the regions with identical {\bmag}.

\begin{figure*}[ht!]
  \centering
  \includegraphics[width=0.7\textwidth]{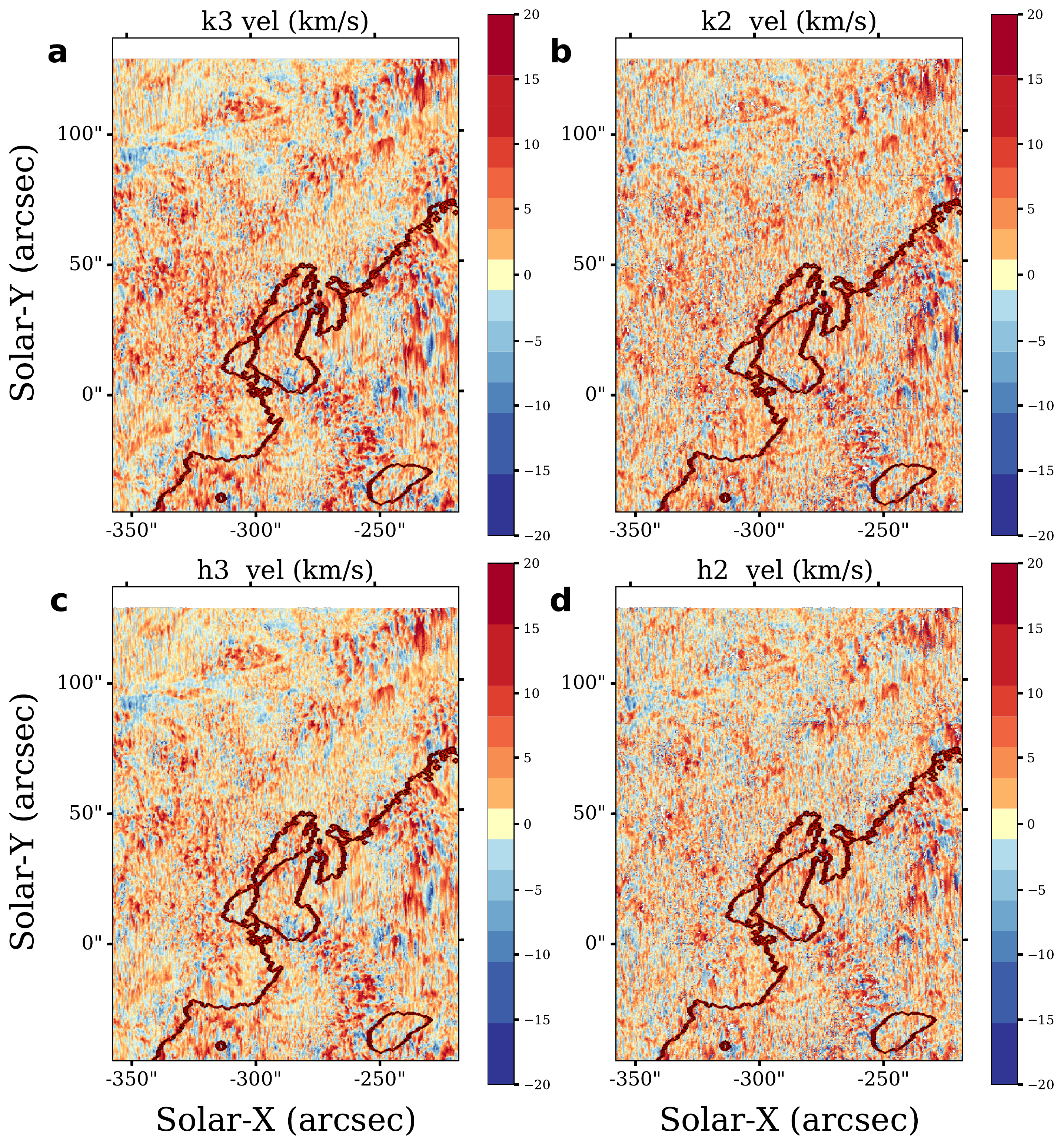}
  \caption{Velocity maps obtained in {\mg} k3 (\textbf{a}), k2 (\textbf{b}), h3 (\textbf{c}) and h2 (\textbf{d}) from DS4. The red contour shows CH-QS boundary. }
  \label{fig:mg2vel_map}
\end{figure*}
\subsubsection{Doppler Shifts} \label{sec:mg2velocity}

To explore and understand the chromospheric velocities further, we now consider the velocities derived from Doppler shifts, which have a tight correlation with local plasma velocity at the height of formation \citep{leenaarts_iris2}. Fig.~\ref{fig:mg2vel_map} displays the velocity maps obtained for k3 (panel a), k2 (panel b), h3 (panel c) and h2 (panel d). Note that while the core velocities are the straightforward shifts from the reference wavelength, the peak velocities are a signed addition of the peak shifts from the reference wavelength. The red contours demarcate CH from QS. The velocity maps for both k and h lines reveal that on average the chromosphere is redshifted in both QS and CH. as observed in {\mg} lines. Moreover, there are no conspicuous differences between CH and QS in the Doppler maps obtained in k3/h3 as well as k2/h2.

\begin{figure*}[hp!]
  \centering
  \includegraphics[width=0.9\textwidth]{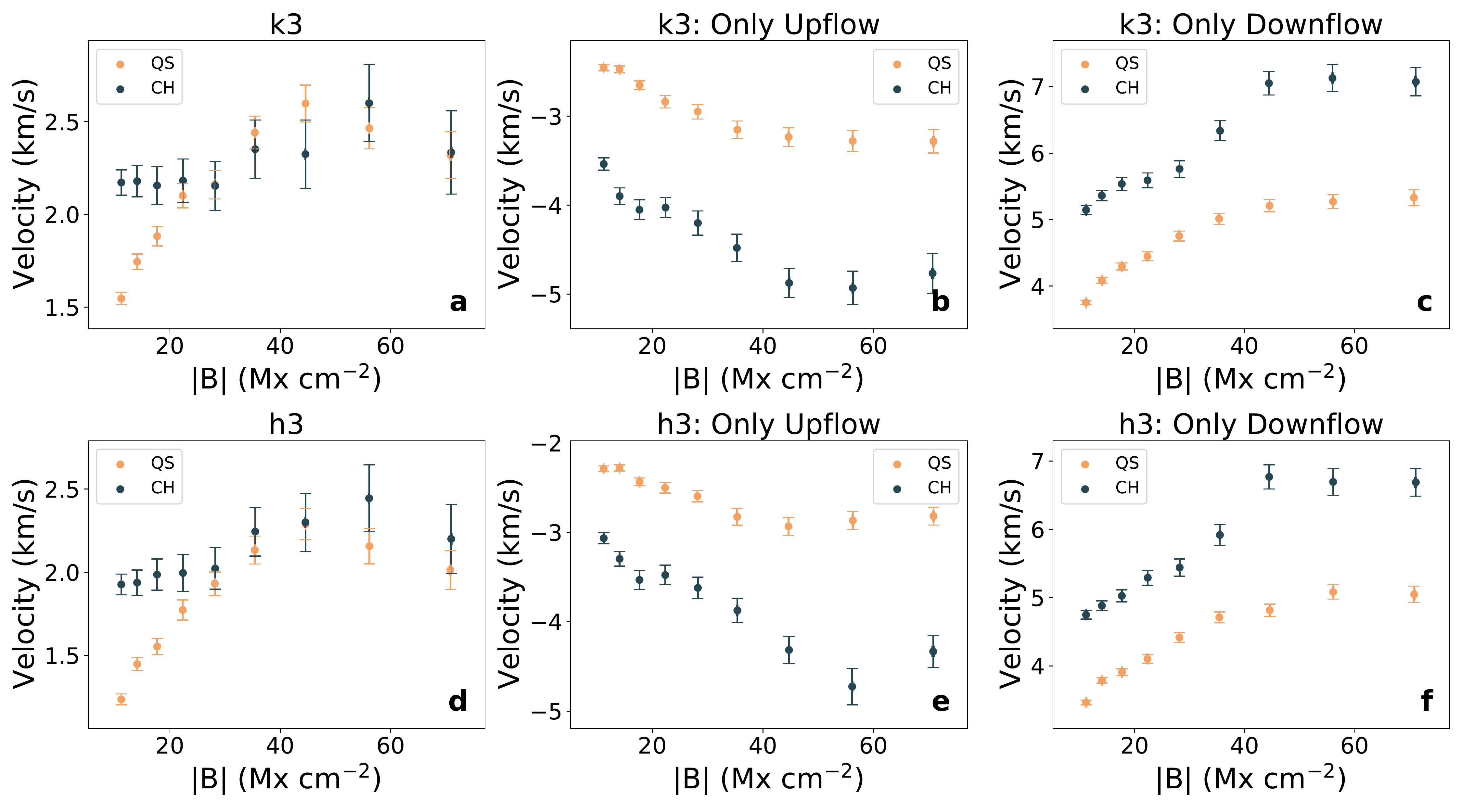}
  \caption{{\mg} k3 and h3 velocity variation with {\bmag}. Panels \textbf{a} and \textbf{d} show the variation of signed average velocities in k3 and h3 binned in {\bmag}. Similarly, panels \textbf{b} and \textbf{e} show the variation of only blueshifted pixels, while panels \textbf{c} and \textbf{f} show the variation of only redshifted pixels. The black (orange) scatter corresponds to CHs (QS).}
  \label{fig:mgcore_vel}
\end{figure*}

In Fig.~\ref{fig:mgcore_vel}, we plot the variation of velocities obtained in k3 (top row) and h3 (bottom row) with {\bmag}. Following {\ppsi} and {\ppc}, we analyze this data in two ways. On the one hand, we consider the signed average velocities in every {\bmag} bin and plot the variation with {\bmag} (panels \textbf{a} \& \textbf{d}). On the other hand, for each bin of {\bmag}, we consider the redshifted and blueshifted pixels separately and plot the variation of velocities with {\bmag} (panels \textbf{b} \& \textbf{e} for upflows and panel \textbf{c} \& \textbf{f} for downflows). While the former provides us the average velocities, the latter gives us a systematic variation of downflows and upflows with increasing {\bmag} in CH and QS. This is akin to the systematic variations seen in Fig.~\ref{fig:mg2peakratio}. Such an exercise can tell us if the dynamics of the magnetic field causes any preferential effect on the redshifts and blueshifts.

\begin{figure*}[hp!]
  \centering
  \includegraphics[width=0.9\textwidth]{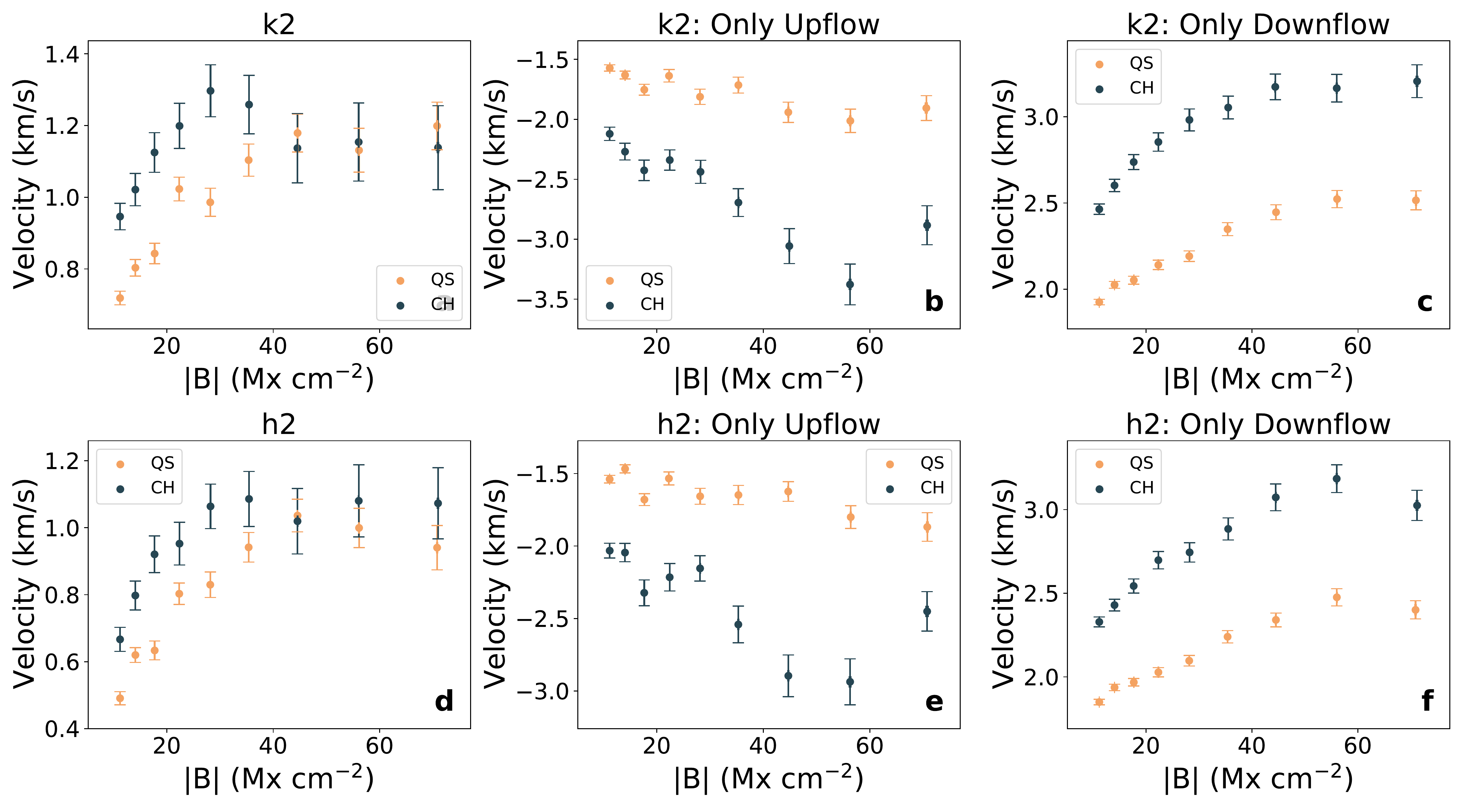}
  \caption{Same as Fig.~\ref{fig:mgcore_vel} but for k2 and h2.}
  \label{fig:mgpeak_vel}
\end{figure*}

Figs.~\ref{fig:mgcore_vel}.\textbf{a} \& \textbf{d} clearly show that, on average, the chromosphere is redshifted in both QS and CH, similar to what is inferred from the maps shown in Fig.~\ref{fig:mg2vel_map}. This result is consistent with the known observations \citep[see e.g.,][and references therein]{Stucki_FUVLinesSumer,Stucki_UVLinesSumer_Chs,Avrett_2013_1DC2}. Moreover, CHs show larger redshift than QS for {\bmag}$\le$30~G, beyond which there are no differences in the velocities. At {\bmag} $\ge$ 80 Gauss, there is some hint for the CHs to show larger redshift. However, note that the average velocities are quite small in both the regions.

When we consider the blue/red shifted pixels separately, both in CH and QS we find a definite increase in the upflow (see Fig.~\ref{fig:mgcore_vel}.\textbf{b} \& \textbf{e}) and downflows (see Fig.~\ref{fig:mgcore_vel}.\textbf{c} \& \textbf{f}) with increasing {\bmag}. Moreover,  the magnitudes of upflows and downflows are larger in CHs than in QS for the regions with identical {\bmag}. Such trend is consistent with the inference made using the ratios of the two peaks shown in Fig.~\ref{fig:mg2peakratio}. Note that the magnitude of the downflows in QS and CH is much larger than that of the upflows, explaining the predominant downflows. Finally, the velocity differences between CHs and QS increase with increasing {\bmag}, with an apparent saturation of velocities for {\bmag} $\ge$ 60 Gauss.

To investigate if these variations are also seen at the average formation height of k2 and h2, we perform the same analysis with the average velocity obtained from the k2 and h2 peaks, and display the results in Fig.~\ref{fig:mgpeak_vel}. The plots (panels \textbf{a} \& \textbf{d}) reveal that the average velocities obtained at k2/h2 peaks are much smaller than the core velocities. The CHs show excess redshifts than QS for regions with {\bmag} $\le$30~G, beyond which the difference in velocities cease to exist. Moreover, the velocities in both CHs and QS increase with {\bmag} till 30~G and saturate thereafter. 

We further note that CHs show excess upflows (\textbf{b} \& \textbf{e}) as well as downflows (\textbf{c} \& \textbf{f} of Fig.~\ref{fig:mgpeak_vel}) over QS for regions with identical {\bmag}. Both upflows and downflows in CHs show a monotonic increase with increasing {\bmag} till about 60~G and saturate thereafter. For QS, however, variation in upflows is very tiny, while downflows do show an increase with increasing {\bmag} that also saturates beyond $\approx$60~G. The velocities obtained from the peaks largely follow the velocities obtained using the core of the line, with the former being smaller than the later.

\subsection{\texorpdfstring{\mg}:: Combined Dataset} \label{mg2_all}
\begin{figure}[t!]
  \centering
\includegraphics[width=\textwidth]{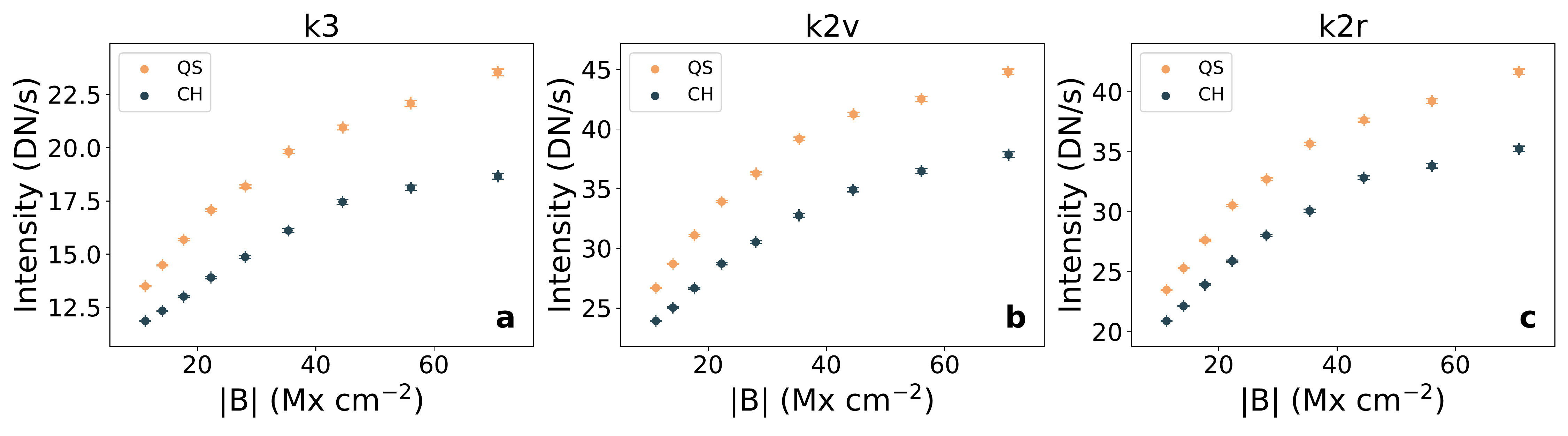}
  \caption{ Same as the top row of Fig.~\ref{fig:mg2intensity}, but for combined dataset. }
  \label{fig:mg2_intens_comb}
\end{figure}
\begin{figure}[ht!]
  \centering
  \includegraphics[width=\textwidth]{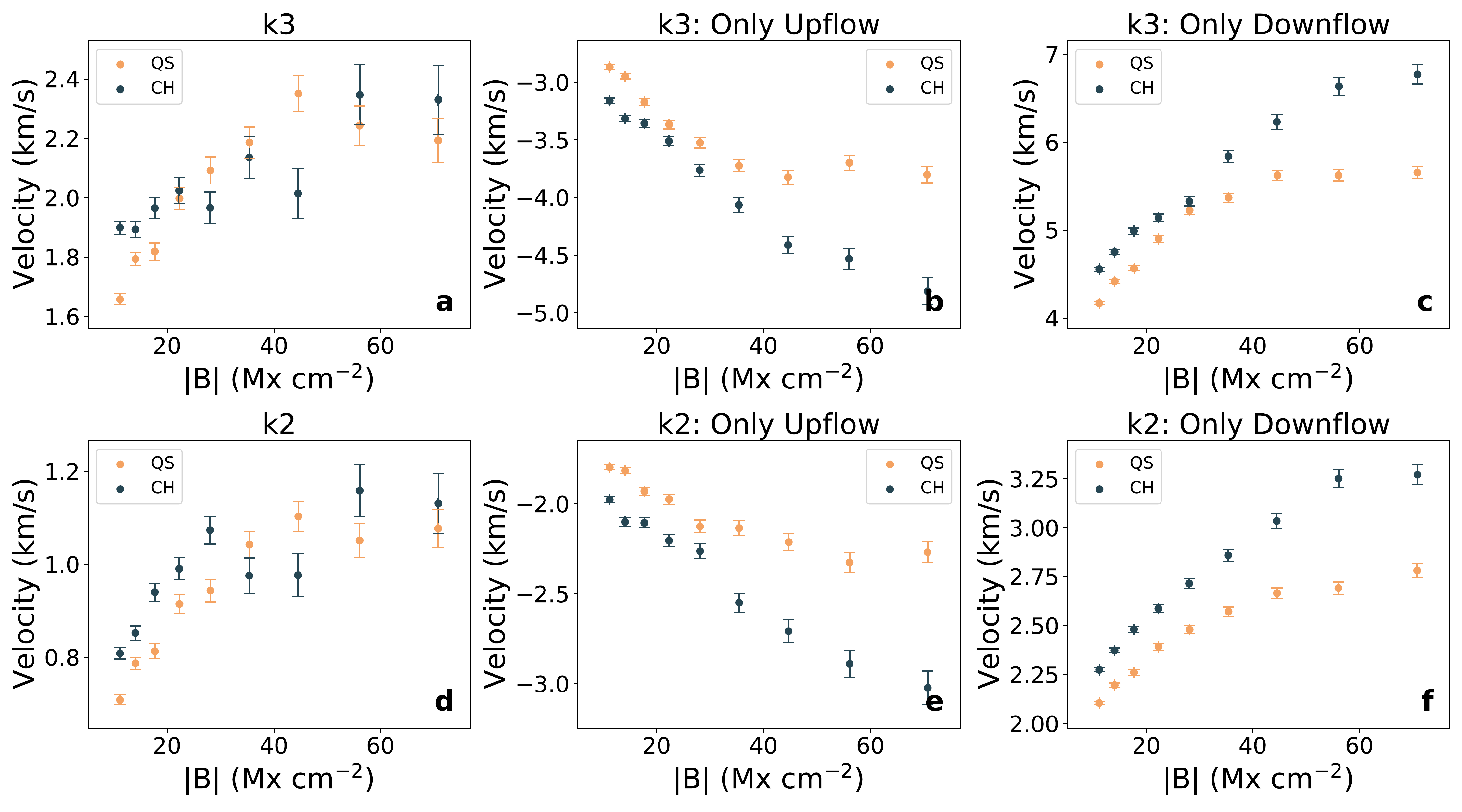}
  \caption{Same as the top rows of Fig.~\ref{fig:mgcore_vel} and Fig.~\ref{fig:mgpeak_vel}, but for the combined dataset.}
  \label{fig:mg2_velocity_comb}
\end{figure}


Having demonstrated the analysis and results obtained for a single dataset, we now consider all the five datasets listed in Table.~\ref{tab:datadetails} to increase the statistical significance of
our results. We emphasize that the results for each dataset is similar to the results reported for DS4 in the previous section. For this purpose, we average the obtained parameters from all five sets of observations, and study the dependence of intensities and velocities on {\bmag}. Note that combining all the dataset is possible because the observations are taken at similar values of $\mu$. We further note that we present the results only for the {\mg}~k line features for brevity as the results for both k and h lines are extremely similar.

In Fig.~\ref{fig:mg2_intens_comb}, we plot the variation of averaged intensities obtained in k3 (panel \textbf{a}), k2v (panel \textbf{b}) and k2r (panel \textbf{c}) as a function of {\bmag}. For all the three features of {\mg}~k line, we find that the intensity increases with increasing {\bmag}, albeit some sign of saturation at higher {\bmag}. We also find that QS regions show excess intensity over CHs for the regions with identical {\bmag} and that the difference in intensities increases with increasing {\bmag}. 

We study the behavior of Doppler shifts as a function of {\bmag} in Fig.~\ref{fig:mg2_velocity_comb}. We plot the signed average of the Doppler shifts of k3 (k2) in Fig.~\ref{fig:mg2_velocity_comb}.\textbf{a} (Fig.~\ref{fig:mg2_velocity_comb}.\textbf{d}). The plots clearly show that both QS and CHs are redshifted on an average, and that the redshift increases with increasing {\bmag}. Moreover, for {\bmag} $\leq$ 30 Gauss, the CHs show marginally excess redshifts, which disappears at higher {\bmag}.  We plot the velocity variation of pixels showing upflows in panels \textbf{b} and \textbf{e} , and of downflows in panels \textbf{c} and \textbf{f}. There is clear signature of monotonic increase of upflows and downflows in CHs with increasing {\bmag}. However such a clear monotonicity is not seen for QS regions. In fact, while the flows do increase for QS till about 30~G, they get saturated thereafter. Moreover, the CHs show larger excess upflows as well as downflows over QS for larger {\bmag}. Finally, the magnitudes of the flows in k3 are larger than that in k2 .

\section{\texorpdfstring{\car}:: Results From the Combined Dataset} \label{sec:c2feature}
\begin{figure}[!ht]
  \centering
  \includegraphics[width=\textwidth]{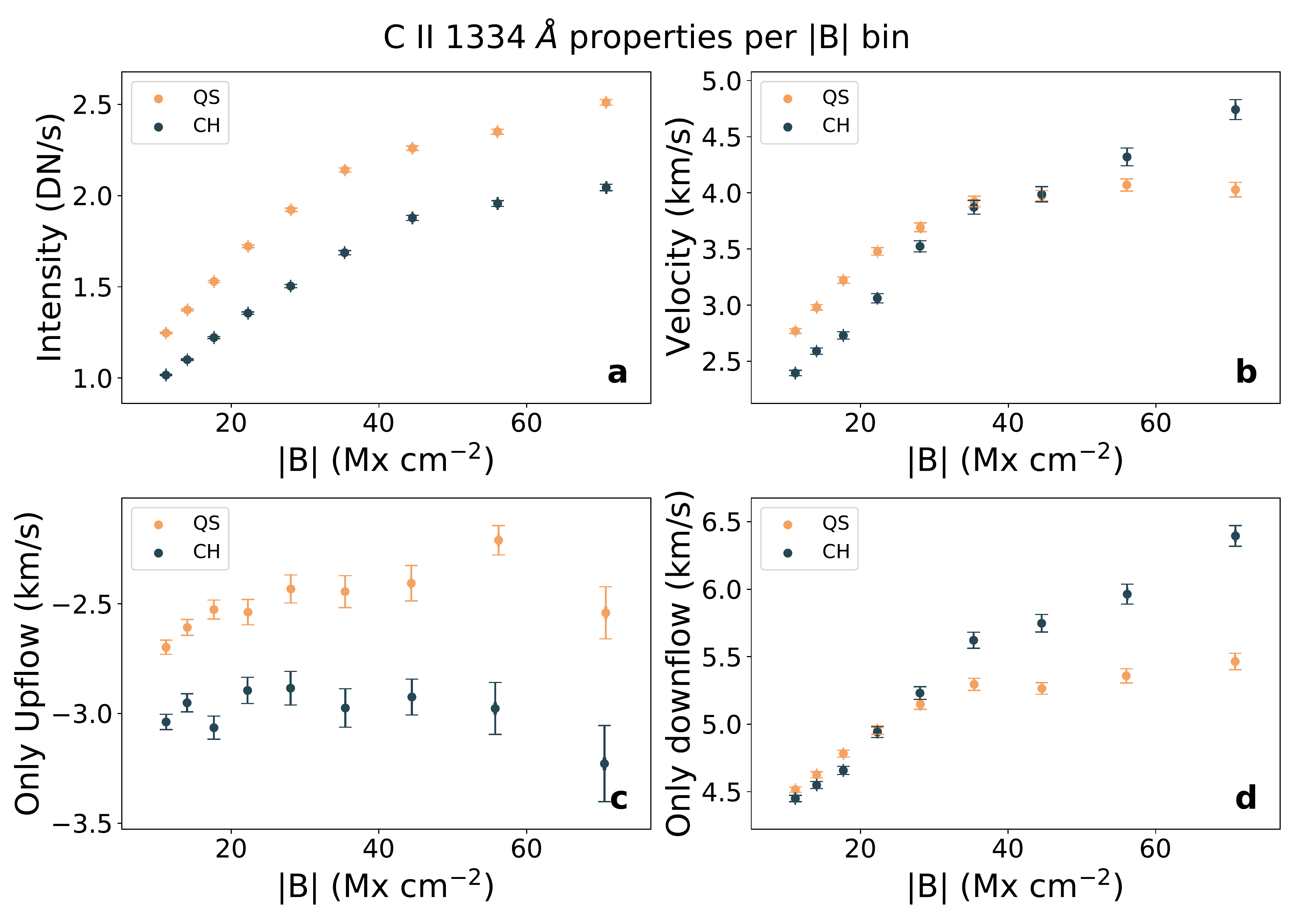}
  \caption{Variation of {\car} 1334{~\AA} properties with {\bmag}. We show the intensity in panel \textbf{a} and velocity panel \textbf{b}. In panel \textbf{c}, only pixels with upflows have been considered, while in panel \textbf{d} only pixels with downflows have been considered. The color scale once again follows black for CH and orange for QS. }
  \label{fig:c2prop}
\end{figure}
The properties of {\car} lines as a function of {\bmag} in CHs and QS using exactly the same five dataset have been studied in great detail in {\ppc}. Here we summarize the salient conclusions from {\ppc}, which are relevant to this paper. The relevant results are graphically summarized in Fig.~\ref{fig:c2prop}, which shows the variation of intensity (panel \textbf{a}), signed average velocity (panel \textbf{b}), average velocity in upflowing (panel \textbf{c}) and downflowing pixels (panel \textbf{d}) as a function of {\bmag}. These variations are by and large similar to the results obtained from the analysis of {\mg} lines in \S\ref{sec:mg}. The intensity, average velocity, upflow and downflow pixels show variation similar to {\mg} with {\bmag}. The differences in CHs and QS in all parameters here  are  similar to the differences seen in {\mg} features.

There are a few notable differences between the properties in {\car} and {\mg} lines, though. The QS shows excess average redshifts for {\bmag}$\le30$ Gauss, while the CHs show excess average redshifts for larger {\bmag} in the {\car} line (see Fig.~\ref{fig:c2prop}.\textbf{b}). Such a relation is not observed for the {\mg} line (see Fig.~\ref{fig:mg2_velocity_comb}). This arises due to consistent redshifts in both CHs and QS for {\bmag}$\le35$ Gauss from Fig.~\ref{fig:c2prop}.\textbf{d}. Note furthermore that the upflow-{\bmag} relation in {\car} is weaker than in {\mg} features.

\section{\texorpdfstring{\si}:: Results From the Combined Dataset}  \label{sec:si4feature}
\begin{figure*}[htpb!]
  \centering
  \includegraphics[width=\textwidth]{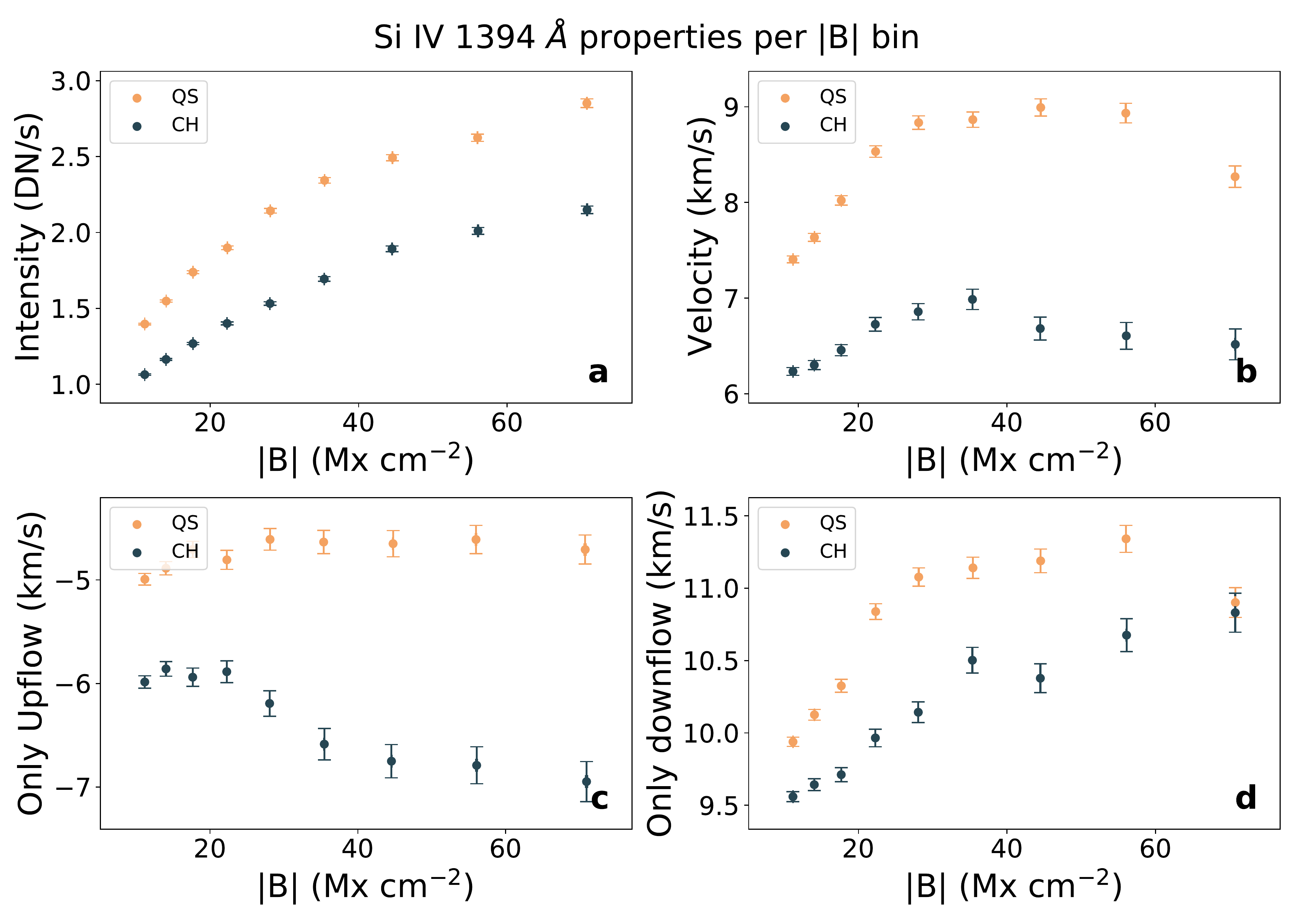}
  \caption{Same as Fig.~\ref{fig:c2prop}, but for {\si} line.}
  \label{fig:si4prop}
\end{figure*}

We now present the results from the intensity and Doppler shift of the {\si} line for all the datasets. The only difference between results presented here and those from {\ppsi} is the inclusion of two additional observations in this work, thereby increasing the statistical significance of the results. We obtain the {\si} line parameters by fitting the spectra with a single Gaussian and constant continuum. The relevant results are graphically summarized in Fig.~\ref{fig:si4prop}. These results are in complete agreement with those obtained in {\ppsi}. Moreover, the results obtained for {\si} bear some similarities with those obtained for {\car} and {\mg} lines. 

The intensity (Fig.~\ref{fig:si4prop}.\textbf{a}) and blueshift (Fig.~\ref{fig:si4prop}.\textbf{c}) differences between CHs and QS, and their relations with {\bmag} are consistent across all three lines, though more enhanced for {\si} line. However, the signed average velocities clearly indicate reduced average redshifts in CHs over QS (Fig.~\ref{fig:si4prop}.\textbf{b}). These average redshifts increase with {\bmag}, and saturate at $\approx40$ Gauss. The CHs show excess blueshifts over QS (Fig.~\ref{fig:si4distr}.\textbf{c}), with the variation similar to those exhibited by the {\car} line (Fig.~\ref{fig:c2prop}.\textbf{c}). The redshifted pixels alone also show a direct relation to {\bmag}, but the QS is more redshifted than CHs (Fig.~\ref{fig:si4prop}.\textbf{d}). Note also that the upflow and downflow velocities obtained for {\si} are much larger than those inferred from {\mg} (Fig.~\ref{fig:mg2_velocity_comb}) or the {\car} lines ({\ppc}). 

\section{Correlations between Doppler shifts of {\mg}, {\car} and {\si}}
\label{sec:combined}

The velocities and intensities show a highly non-trivial relation as a function of the formation height of different spectral lines. Therefore, to investigate if there is any correlation between Doppler signatures observed in the three different spectral lines {viz.} {\mg}, {\car} and {\si}, we consider the approximate formation height of these lines obtained from numerical simulations. It has been suggested that on average {\car} lines form slightly higher in the atmosphere than the {\mg} lines. Within the {\mg} lines, the k line forms higher than the h line. Moreover, it has also been found that the line cores of both k \& h lines forms higher than their respective peaks \citep{leenaarts_iris2,Rathore_CII_paper2}. The {\si} line forms in optically thin conditions, so ascribing an exact formation height is not possible. However, it forms at a higher temperature in the TR. We may, therefore, ascribe a greater height to {\si} than the {\mg} and {\car} lines. With this prior, we may assume that the formation height (ascending order) is approximately {\mg} h2 $\le$ {\mg} k2 $<$ {\mg} h3 $\le$ {\mg} k3 $\approx$ {\car} $<$ {\si}.

The obtained velocities in different line features of {\mg} (Fig.~\ref{fig:mg2_velocity_comb}. \textbf{b}, \textbf{c}, \textbf{e} \& \textbf{f}), {\car} (Fig.\ref{fig:c2prop}. \textbf{c} \& \textbf{d}) and {\si} (Fig.~\ref{fig:si4prop}. \textbf{c} \& \textbf{d}) clearly show that the velocity magnitude increases with increasing formation height. Considering mass flux conserving flows~\citep{Avrett_2013_1DC2}, and that the density decreases as a function of height in the solar atmosphere, it is plausible to hypothesize that the upflows (downflows) at lower (greater) heights are enhanced (reduced) while traveling towards greater (lower) heights.

To check this hypothesis, we investigate the correlations between {\mg}, {\car} and {\si} velocities. Since {\mg} and {\car} form at approximately the same height, we expect these two lines to have similar properties vis-\`a-vis the {\si} line. In {\ppsi}, it is suggested that the increase in {\si} blueshift with increasing with {\bmag} may indicate the signatures of the solar wind emergence. This motivates us to explore if the observed flows in chromosphere detected in {\mg} and {\car} lines are in any way related to those obtained from {\si}. For this analysis, we use the results obtained from the combined dataset. Note that we only consider the {\mg} k2, {\mg} k3, {\car}~1334~{\AA} and {\si} lines in this analysis. We emphasize that the results from {\mg} h follow the results from k feature and hence are not shown for brevity.

\begin{figure*}[!htpb]
  \includegraphics[width=\linewidth]{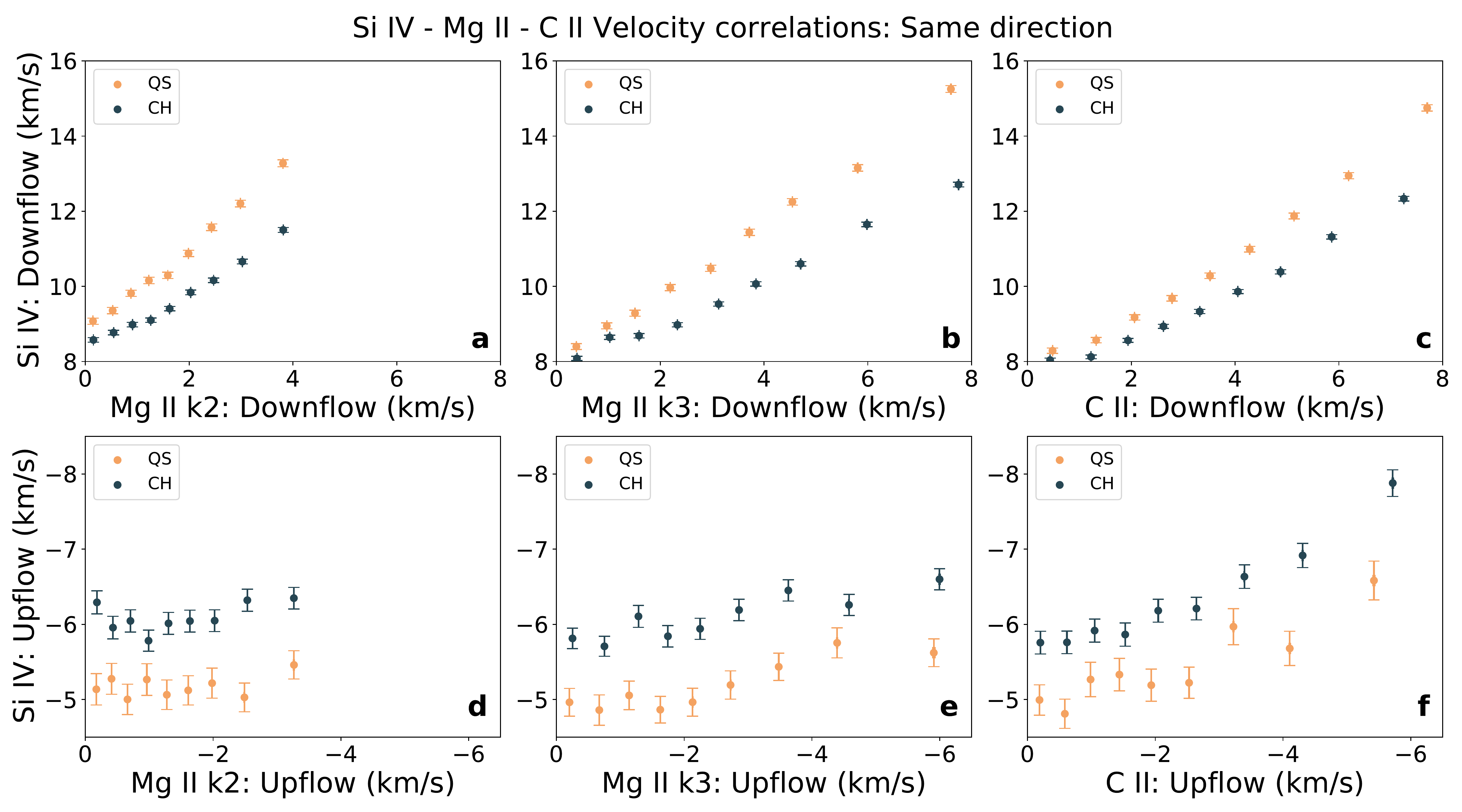}
  \caption{Inter correlations between {{\si}} and {\mg} k2, {\mg} k3, {\car} Doppler shifts. The top row depicts correlations between the downflows in {\mg}, {\car} and {\si}, while the bottom row depicts correlations between upflows in {\mg}, {\car}, and {\si}. The columns follow approximate formation height from {\mg} k2 to {\car}.}
  \label{fig:SiMgC_same}
\end{figure*}
\begin{figure*}[!htpb]
  \includegraphics[width=\linewidth]{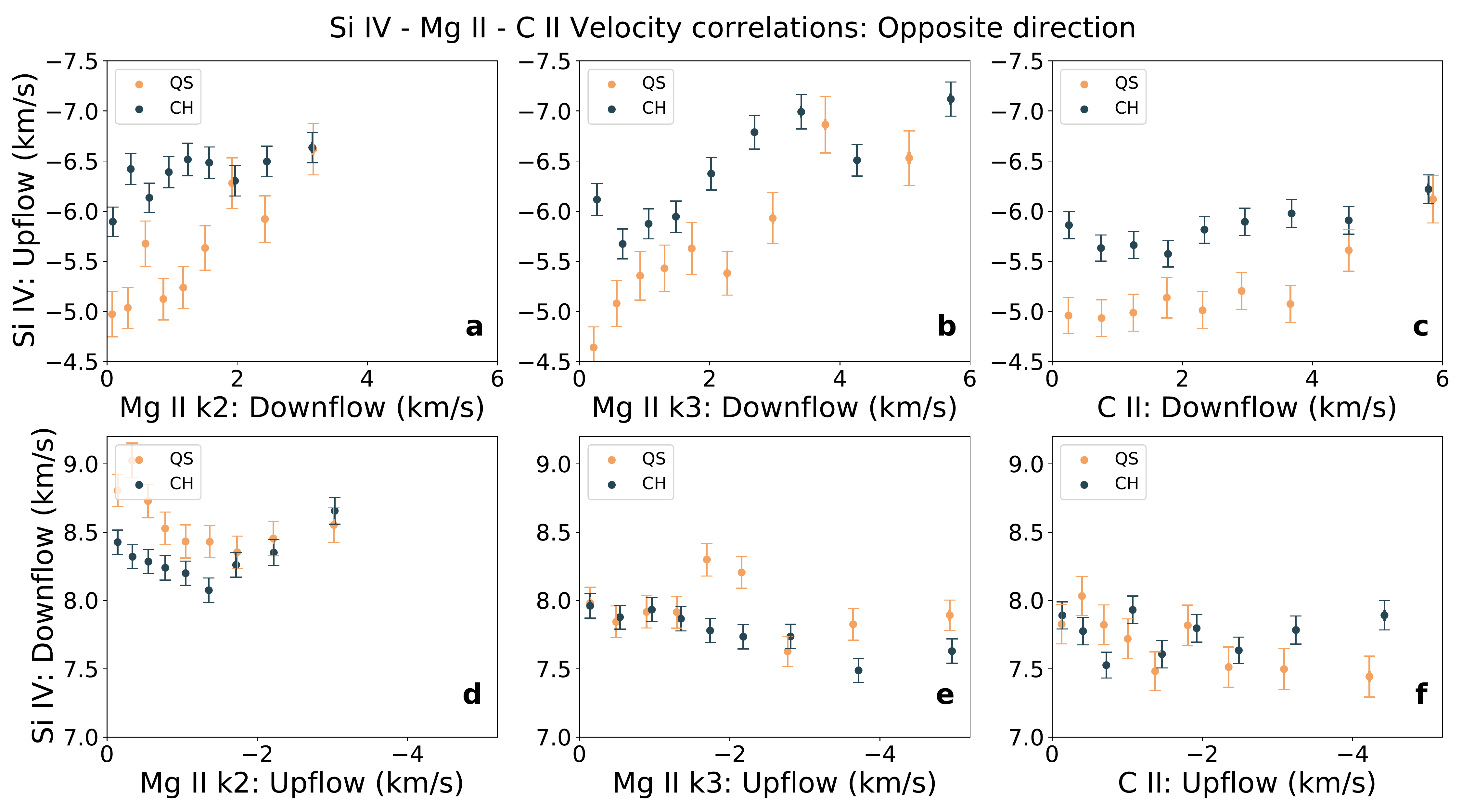}
  \caption{Inter correlations between {\si} and {\mg} k2, {\mg} k3, {\car} Doppler shifts. The top row depicts correlations between the downflows in {\mg} \& {\car} with upflows in {\si}, while the bottom row depicts correlations between upflows in {\mg} \& {\car} with downflows {\si}. The columns follow approximate formation height from {\mg} k2 to {\car}. }
  \label{fig:SiMgC_opp}
\end{figure*}

We split the velocities observed in {\mg}, {\car} and {\si} into sets of pixels containing upflows and downflows. Then, we consider scatter plots between flows in intersection of these sets, e.g., relation between pixels showing upflows in {\mg} k3 \& upflows in {\si}, upflows in {\mg} k3 \& downflows in {\si} and so on. These scatter plots are obtained for {\si} velocities, in {\mg} and {\car} velocity bins to improve statistics. Note that the bins here are selected in deciles, i.e, every 10\% of the data for each {\mg} or {\car} feature is considered to be in one bin.

In Fig.~\ref{fig:SiMgC_same}, we plot the correlations between downflows (top row) and upflows (bottom row) observed in {\si} with those observed in {\mg}~k2 (panels \textbf{a} \& \textbf{d}), {\mg}~k3 (panels \textbf{b} \& \textbf{e}) and {\car} (panels \textbf{c} and \textbf{f}). Panels \textbf{a}, \textbf{b} and \textbf{c} demonstrate that the downflows observed in {\si} are strongly correlated with those observed in {\mg} k2, k3 and {\car}. For a given value of downflow in {\si}, the downflows are stronger in k3 and {\car} than those in k2. Note though, that the downflows in k3 and {\car} are very similar. Moreover, {\si} displays excess downflows in QS vis-\`a-vis CH for similar {\car} and {\mg} downflows. These differences in QS-CH {\si} downflows are also observed to increase with increasing {\mg} and {\car} downflows. Similarly, Panels \textbf{d}, \textbf{e} and \textbf{f} suggest that the upflows in {\si} have slightly better correlation with upflows in {\mg} k3 and {\car} than those in {\mg} k2. Moreover, like downflows, we find that for the similar upflows in {\mg} and {\car}, the CH exhibit larger upflows vis-\`a-vis QS in {\si}. We further note that there is a slight hint of increase in the difference of upflows observed in CH and QS in {\mg} and {\car} lines.

In Fig.~\ref{fig:SiMgC_opp}, we study the correlations between the upflows in {\si} with downflows in {\mg} and {\car} and vice-versa as shown in the top and bottom rows, respectively. We find that the upflows in {\si} have a monotonic relation with the downflows in {\mg} and {\car} (top row). In addition, we note that for similar downflows observed in {\mg} and {\car}, {\si} shows stronger upflows in CH and than in QS. On the other hand, the {\si} downflows do not show any particular correlation with {\mg} and {\car} upflows. Furthermore, the {\si} downflows in CH and QS remain consistent for similar upflows in {\mg} and {\car}. We do not find any relation between the downflows in {\si} with upflows in {\mg} and {\car}.

\section{Inferences from intensity and velocity diagnostics} \label{summary}
The problems of coronal heating in QS and CHs, and the formation and acceleration of solar wind are intimately tied to the structure and dynamics of the magnetic field in the respective regions. Therefore, comparative studies between CHs and QS become important in understanding the plasma dynamics, and the underlying processes. The {\mg}, {\car} and {\si} lines observed by IRIS probe different layers in the chromosphere \& TR, have provided us with a unique opportunity to understand the dynamics of these regions as a dynamically coupled system. In this paper, we characterize, in detail, the dynamics of the {\mg} line by combining the information related to the plasma dynamics with that of the magnetic field. In addition, we also probe the correlations between the Doppler shift obtained for {\mg} lines, and those obtained for the {\si} line from {\ppsi} and the {\car} line from {\ppc}, to investigate if a common origin may be ascribed to the observed dynamics in the different lines. 
Below we summarize and discuss our results followed by an interpretation in \S\ref{interpretation}.


\subsection{Intensity differences}\label{intensity}
The intensities in the {\mg} h, k lines (both in the core and peaks) \& {\car} lines formed in the chromosphere, {\si} line formed in the TR increase with {\bmag} (see Fig.~\ref{fig:mg2_intens_comb}, Fig.~\ref{fig:c2prop}.a and  Fig.~\ref{fig:si4prop}.a). For all the three lines, CHs show reduced intensity over QS for the regions with identical {\bmag}. Moreover, the difference in the intensities increases with increasing {\bmag}. The observed differences in the chromospheric intensities in CHs and QS suggest that CHs have lower source function values over QS for regions with identical {\bmag}~\citep{Rathore_CII_paper2}.

Our results further show that the differences between CH and QS intensities exist already at the chromospheric level for a given magnetic flux region \citep[see also][\& {\ppc}]{PradeepKashyap2018}. We note that the ratios of QS to CH intensities in the largest {\bmag} ($\approx80$ Gauss) bins are smallest in {\mg}~k2, and increase through {\car}, {\mg}~k3 and {\si} lines as 1.18, 1.22, 1.26, and 1.32, respectively, suggesting an increasing differentiation between CH and QS from low chromosphere to TR. 

The intensity differences in the CHs and QS in the corona are well known \citep[see, e.g.,][]{Krieger1973}. However, such differences are not seen in either chromospheric or TR above noise level \citep[][]{Stucki_UVLinesSumer_Chs,xia_chsumer}. These observations led \cite{Weigelmann_loopstats} to attribute the CH-QS intensity differences to loop statistics in these regions. Based on potential field extrapolations, \cite{Weigelmann_loopstats} found that the QS have excess number of longer closed loops over CH, while similar numbers of shorter closed loops are present in both the regions. Using the scaling laws, valid for optically thin plasma, they proposed that the reduction in CH intensity over QS naturally comes out as a function of deficit of longer loops in CHs.
    
While the scenario proposed by \cite{Weigelmann_loopstats} is used to explain the intensity difference in {\si} in {\ppsi}, it may not be directly applicable to the chromosphere, as also argued by \cite{PradeepKashyap2018}. However, since the loop statistics of \cite{Weigelmann_loopstats} is derived from the extrapolations of photospheric magnetograms, it may be plausible to suggest that the statistics itself (and not the scaling relation for plasma emission) is also valid for the chromosphere. Therefore, we may conclude that at a relatively higher {\bmag} a deficit of shorter loops in the CHs with respect to the QS is observed already in the low chromosphere. The source function of the chromospheric lines may in-part be influenced by this loop statistics at chromospheric heights and may explain the marginal deficit of intensities in CHs over QS.

\subsection{Doppler shift: Variations and correlations}\label{flows}


Doppler measurements in all three lines demonstrate that on average both the chromosphere \& TR are red-shifted (see. Fig.~\ref{fig:mg2_velocity_comb}.a \&~d, Fig.~\ref{fig:c2prop}.b, and Fig.~\ref{fig:si4prop}.b). The average redshifts are found to increase with {\bmag}, and increase from {\mg}~k2 to {\si} for similar {\bmag}. By studying the red-shifted and blue-shifted pixels separately, we find that in the chromospheric lines, CHs have excess upflows as well as downflows vis-\'a-vis QS for identical {\bmag} and that the excess increases with increasing {\bmag} (see Fig.~\ref{fig:mg2_velocity_comb}, and Fig.~\ref{fig:c2prop}.c \& d). However, in the TR, CHs have excess (reduced) upflows (downflows) over QS for the regions with identical {\bmag} (see Fig.~\ref{fig:si4prop}. c \& d). With uncertainties, the magnitudes of upflows and downflows are in an approximate descending order of {\si}, {\mg} cores, {\car} and {\mg} peak. We further note that, while {\mg}~k3 and {\car} lines show similar downflows, the upflows are larger in {\mg}~k3.

To asses any (or otherwise) association between the flows observed in the chromosphere and TR, we perform a correlation study in intersection of sets of pixels which show flows in different lines. That is, we study mean variation of upflows in the TR pixels which also show upflows in the chromosphere, and so on for different combinations of flows. This analysis gives us the variation of mean TR flows with chromospheric flows, and provides information on the persistence of flows in different lines. We find that the flows in chromosphere and TR are tightly correlated, i.e., the downflows in chromosphere with both upflows and downflows in TR, upflows in chromosphere with those in TR. However, we did not find any correlation between chromospheric upflows with downflows in TR. Moreover, for the similar downflows (upflows) in the chromosphere the CHs show reduced TR downflows (excess upflows) over QS. Additionally, for similar downflows in the chromosphere, the CHs show excess upflows over QS in the TR.

The observations reported here lead to two questions. Firstly, what physical mechanism(s) give rise to these flows? Secondly, is it possible to explain the observed differences between the flows observed in CH and QS in the chromosphere and TR, including the difference in the intensities discussed in \S\ref{intensity}? While we deal with the former here, later is taken up in the \S\ref{interpretation}.

The tight correlations between TR downflows measured using {\si} and those observed in the chromosphere measured using {\mg} and {\car} may either be explained by field aligned downflows due to condensations from corona to TR to chromosphere \citep[see, e.g.,][]{klimchuk_2006_coronalheating, TriMD_2009,TriMK_2010,TriMK_2012} or due to return flows of type-\rm{II} spicules \citep[][]{Kli_2012, GhoKT_2019, GhoTK_2021, BosRJ_2021}. However, we note that the observed magnitude of the TR downflows are much larger than those predicted using 1D hydrodynamic simulations of coronal impulsive heating followed by evaporation and condensation. Therefore, for the reasons elaborated in \cite{GhoKT_2019, GhoTK_2021}, it is more likely that the observed downflows here are due to the return flows of type~\rm{II} spicules. Our finding that the speeds of chromospheric downflows is lower than those in TR is very likely due to the plasma flowing from lower density to higher density. Note, however, that the net deceleration of the plasma is dependent on the interplay of deceleration due to atmospheric stratification \& magnetic pressure, and acceleration due to gravity \& plasma compression. 

Our observations further show that the upflows in chromosphere and TR are also correlated. Moreover, these upflows show an increase in magnitude with increasing {\bmag} as well as atmospheric height. This may be possible if the upflows are moving through an expanding flux tube, under the assumption of constant mass flux. The upflows themselves, however, may have been caused by launch of events like Type~\rm{II} spicules \citep{Bart_2007_TwoSpicules,Tian_2008_MassSuply,Tian_2014_smallscalejets,Tanmoy_2019_spicules}. However, note that such upflows may also be generated due to upward propagating waves \citep[e.g.,][]{cranmer_2005_AlvenWave}. Since Alfv\'en waves are known to be ubiquitous in the chromosphere~\citep{Bart_2007_Alfvenwaves}, disentangling the exact effects of Alfv\'en wave v/s spicule-like propagation upward is difficult \citep[see, however,][]{GhoKT_2019,GhoTK_2021}.

Finally, the chromospheric downflows also bear a direct relationship with the upflows in TR. Such correlations suggest a common origin of these flows and hints towards existence of bidirectional flows. Bidirectional flows have been observed in QS \& CH (predominantly occurring in the CH) network regions by \cite{Aiouaz_2008_bidirectionalflows}, and in active regions by \cite{Barczynski_2021_ARUpflowMg2downflows} as redshift in TR and blueshifts in the low corona \citep[see also][for bidirectional flows in transient events]{GupST_2018}. We propose that such bidirectional flows occurring between the chromosphere and the TR can suitably explain our observations

\begin{figure*}[ht!]
    \centering
    \includegraphics[width=\linewidth]{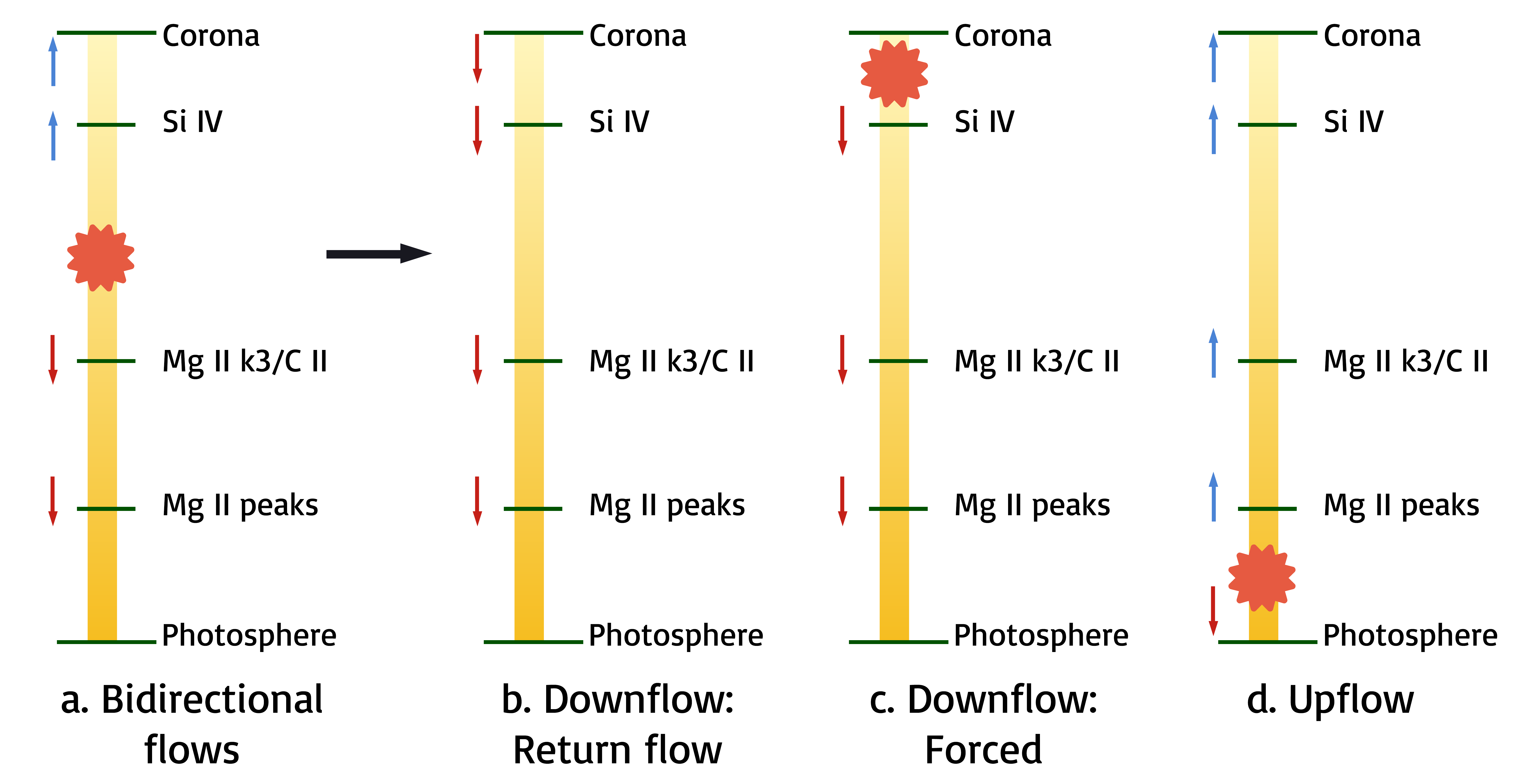}
    \caption{A schematic depicting a unified picture of flow generation, including observed correlations between flows. The vertical bar denotes density reducing with height from dark yellow to white. The red asterisk depicts an impulsive event, which gives rise to flows (arrows). Blue upward arrows depict upflows, and red downward arrows depict downflows.  Panel {\bf a} shows the basic bidirectional flow generation, which eventually gives rise to the return flow in Panel {\bf b}. Impulsive events occurring much higher than {\si} formation height, giving rise to downflows is shown in Panel {\bf c}. Upflows generated due to impulsive events, low in the atmosphere are depicted in Panel {\bf d}. See in-text for details.}
    \label{fig:flow_relations}
\end{figure*}

The scenario we propose is illustrated in Fig.~\ref{fig:flow_relations}. The vertical color bars changing from deep yellow in the photosphere to white in the corona indicate reducing density with increasing height in the atmosphere. The approximate formation height (or rather, a proxy for temperature) of different ions corresponding to the spectral lines studied here are labeled with horizontal lines. The `asterisks' indicate the location of an impulsive event, while the arrows mark the direction of expected flows. The blue (red) arrows indicating upflows (downflows). 

We present four different scenarios, based on the same physical mechanism, to explain the three set of observation. As evidenced by similar skew \& kurtosis in {\car} 1334~{\AA} line (see {\ppc}) and non thermal widths in {\si} (see {\ppsi}), it is plausible to conclude that similar physical mechanisms give rise to the observed spectral profiles in CHs and QS. This mechanism, in our interpretation, is an impulsive dumping of energy in CHs and QS. For an impulsive event occurring between the formation height of {\si} and {\car} or {\mg}, bi-directional flows will be produced in form of upflows in {\si} and downflows in {\car} and {\mg}. Since the chromosphere is denser than the TR, the chromospheric radiative cooling time scales are smaller. Hence, the downflows would cool down faster, and be visible in cooler lines like {\mg} and {\car}, while the upflows persist in relatively hotter lines like {\si}. Some of the upflows observed in {\si} may persist till greater heights and then fall back, similar to Type~\rm{II} spicule return flows (Fig.~\ref{fig:flow_relations}.b). The returning flows will be observed as persistent downflows in all three lines, with descending speeds. We note, however, that the persistent downflows may also be caused by the impulsive event occurring above or about the formation height of {\si} (Fig.~\ref{fig:flow_relations}.c), and may have similar signatures of descending speeds as in previous case. Finally, the impulsive events may also occur either below or at the height of formation of {\mg} peaks (Fig.~\ref{fig:flow_relations}.d) resulting into launch of chromospheric jets that may show persistent upflows in chromosphere and TR, followed by the downflows at later times.

Bidirectional flows in an expanding, cylindrically symmetric flux tube have been observed in field aligned 1-D simulations by \cite{He2008_SWModelling}. In these simulations, impulsive events deposit energy at the height of $\approx5$~Mm, which is also the location where expansion of the flux tube starts. The results demonstrate that at the onset of the impulsive event the plasma moves outward from location of impulse, showing bidirectional flows. \cite{He2008_SWModelling} show that the velocities in \ion{Si}{2}, \ion{C}{4} and \ion{Ne}{8} obtained from their setup match those observed by \cite{tu2005solar}. \cite{He2008_SWModelling} further show that the downflows at $10^4$ K are $\approx2$ km/s. The magnitude of these downflows increase with height, reaching up-to $4$~km/s near the energy dumping heights. These velocities were obtained with a {\bmag} of $56.5$ Gauss in the simulation. The velocities obtained in this simulation are consistent with the downflow speeds observed in {\mg} k2 (see Fig.~\ref{fig:mg2_velocity_comb}.f) at $\approx56$ Gauss, but it is much lower than the downflows speeds observed in {\mg} k3 and {\car}.

\cite{Hansteen2010_SWModelling} perform a 3D simulation of a QS region for different average {\bmag}, spanning from the convection zone to the corona. 
It is found that impulsive events due to reconnection occurring at various heights, give rise to bi-directional flows, seen as co-spatial blueshifts (redshifts) in the corona (TR) concentrated at loop footpoints.

Such events occurring across a range of heights can give rise to correlated flows similar to the results reported here. \cite{Hansteen2010_SWModelling}, in the their B1 model set up, were able to reproduce velocities consistent with blueshifts observed in the corona \citep[see Fig. 11 of][]{Hansteen2010_SWModelling}. Note however, that the downflows speeds near the formation temperature of {\mg} and {\car} inferred from these simulations were between $2-4$ km/s, which are lesser than those reported in this work. Nevertheless, we note that the scenario presented by \cite{Hansteen2010_SWModelling} may potentially explain the correlated bidirectional flows reported in this paper. 

Finally, while the above described scenario based on impulsive events may explain the observed downflows, it is important to highlight that spicule like flows may also be obtained due to `squeezing' of flux tubes near the chromosphere \citep[see e.g.][and also \citet{IsoTA_2007} ]{Bart_2007_TwoSpicules,Juan_2011_spiculesim,Juan_2017_spicules,Juan_2019_spiculelikeflow}. The rising plasma from the 'squeeze' has been observed to be heated up and detected in various IRIS lines such as {\mg} and {\si} \citep[][]{Juan_2017_spicules}. Thus, the persistent upflows may also be explained through such spicule-like flows, while the downflows may then be explained by the return of such spicule like flows. Note that, throughout the paper, we assume that the type-II spicules are produced due to impulsive events~\citep[][]{Moore_2011_xrayjets,Moore_2013_jetrotation,Juan_2011_spiculesim,Tanmoy_2019_spicules}.  
\section{A unified scenario the origin of solar wind, switchbacks and QS heating}\label{interpretation}
While the occurrence of impulsive events at the interface between chromosphere and corona may explain the observed flow variations with {\bmag} in different lines and their interrelations, the question remains as to what leads to the observed differences between the intensities as well as flows in CHs and QS. To explain the differences in the intensities, in \S\ref{intensity} we invoked the loop statistics in CHs and QS derived by \cite{Weigelmann_loopstats}. The predominant velocity differences between CH and QS are: i) reduced {\si} downflows in CHs over QS for similar downflows in the chromospheric lines, and ii) excess {\si} upflows in CHs over QS for similar upflows and downflows in the chromospheric lines. These results indicate excess acceleration of upflows in CHs and excess deceleration of downflows in QS. Furthermore, while the QS shows enhanced intensity over CHs for regions with similar {\bmag}, the CHs show larger flow speeds (except {\si} downflows) over QS. Such an observation thus hints towards a unified scenario of heating of the corona in QS and CH as well as the emergence of the solar wind. Therefore, we then ask if it is possible to combine the loop statistics and the occurrence of flows due to impulsive events illustrated in Fig.~\ref{fig:flow_relations}, to explain the observed differences in intensities as well as the Doppler shifts in CHs and QS, similar to that is discussed in \cite{TriNS_2021} for {\si}. 

\begin{figure*}[htpb!]
    \centering
  \includegraphics[width=0.49\textwidth]{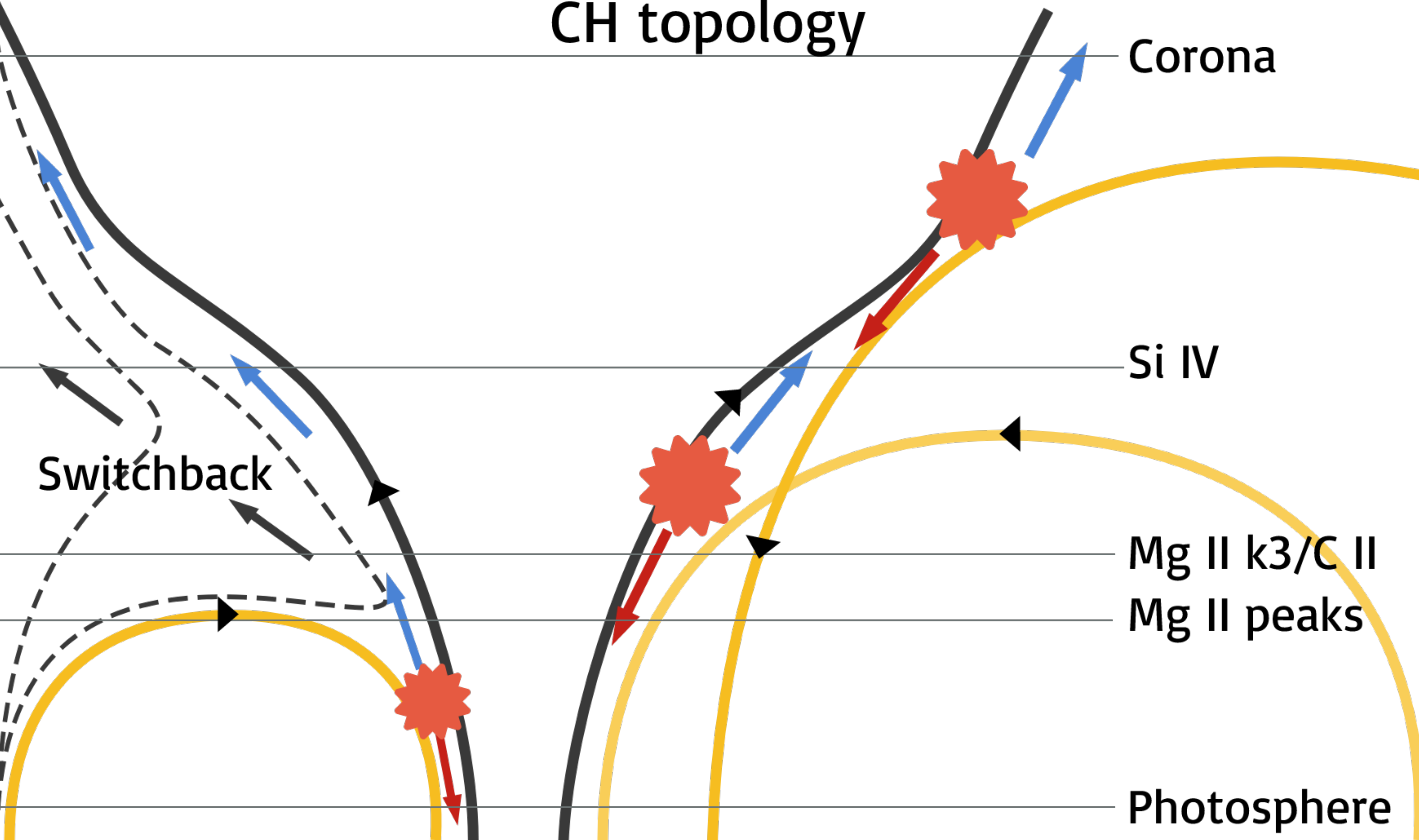}
  \includegraphics[width=0.49\textwidth]{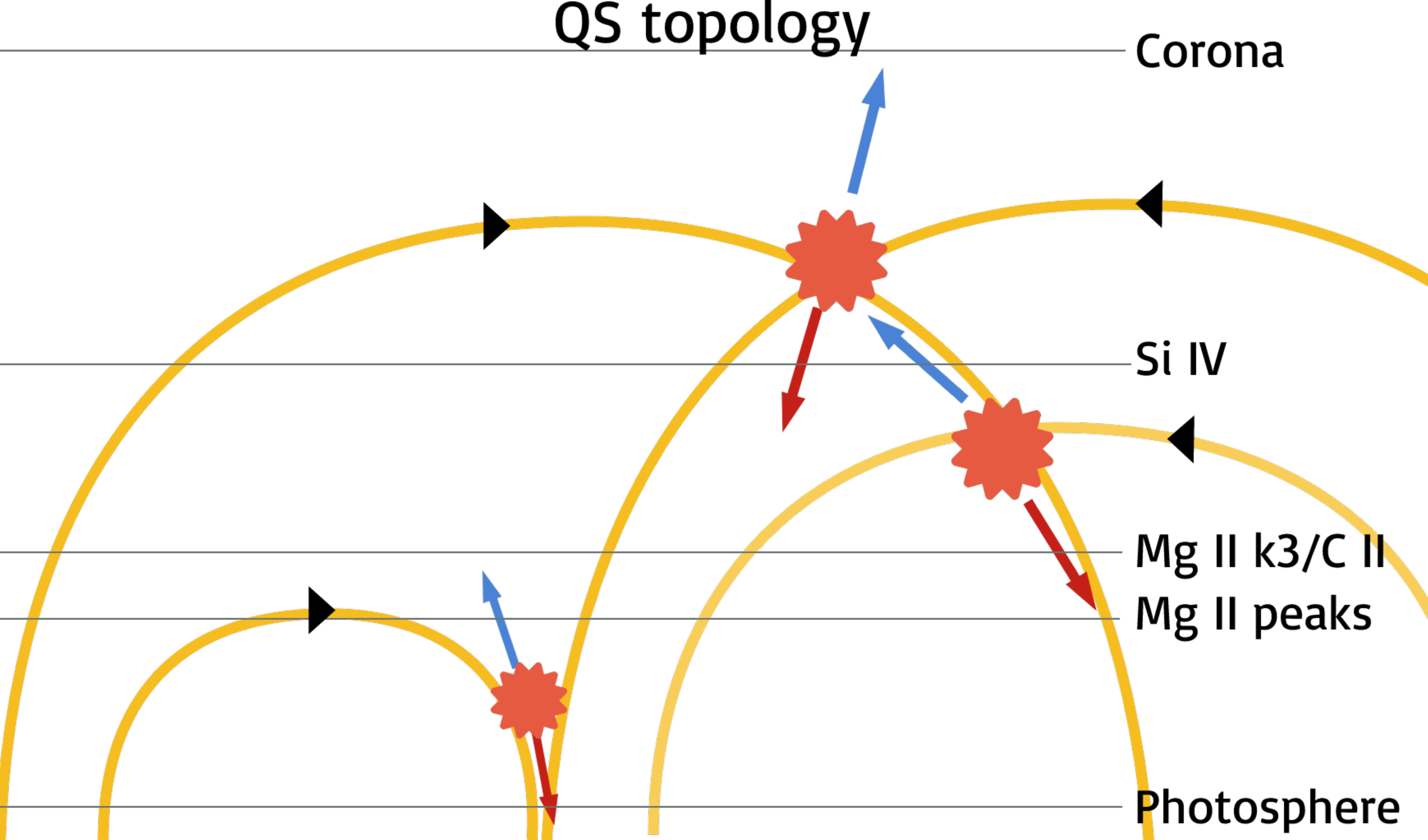}
  \caption{A schematic depicting the proposed picture of impulsive heating occurring across different magnetic field topologies. In the left panel, we show a CH topology, including open funnel like structure (black) and closed loops of varying sizes (yellow) Impulsive events (red asterisks) due to interchange reconnection between the open and closed field lines give rise to bidirectional flows (blue and red arrows). Of these flows, the upflows are enhanced due to density stratification and the expanding flux tube in CHs. Interchange reconnection may occur over a range of heights, and the corresponding bidirectional flows may be observed across different spectral lines marked in approximate order of formation heights. An example of the reconnected field line propagating outward as a switchback is depicted as dashed line, with the approximate propagation direction depicted by black arrows. Right panel: QS topology depicted with the same terminology as CH topology. Note here that while one does expect correlated bidirectional flows in QS, the upflows are not accelerated due to the absence of funnel-like structures. See in-text for more details. }
  \label{fig:SiC}
\end{figure*}
A graphic depicting the scenario we propose is shown in Fig.~\ref{fig:SiC}. The left panel depicts the predominant topology in CHs while the right panel is for QS regions, based on the loop statistics of \cite{Weigelmann_loopstats}, according to which both CHs and QS have equal number of short closed loops but QS has predominantly large closed loops and CHs have open field lines. In CH regions, the interchange reconnection \citep[e.g.,][]{Fis_2005,JanTM_2008}, leading to impulsive events, may occur between closed and open field lines, while in QS, the impulsive events will be due to reconnection among closed-closed loops, \vishal{similar to those observed in the core of active regions during transient formation of loops \citep[see][]{Tri_2021}}. The excess open, and expanding flux tubes in CHs may cause preferential acceleration of upflows in CHs over QS. In principle, the scenario proposed here is similar to those employed by \cite{Tian_2008_EmissionlinesPFSS,Tian_2008_MassSuply,He_2007_SWQS} to explain the Doppler shifts observed in QS-CH in coronal and TR lines, and similar to the Fig.~5 of \cite{He_2010_solarwindorigins}. Note that the concept of interchange reconnection has been invoked to explain active region outflows by \cite{DelZanna_2011_interchangereconnect,Barczynski_2021_ARUpflowMg2downflows}, solar wind disappearance events by \cite{JanTM_2008}, active region jets\vishal{, X-ray \& cool jets in polar coronal holes~\citep{Moore_2011_xrayjets,Moore_2013_jetrotation,Moore_2015_untwist_CH}} and type III radio burst \citep[e.g.,][]{2016_sargam_type3burst}. 

Under the scenario presented in Fig.~\ref{fig:SiC}, the impulsive events may occur across a range of heights via magnetic reconnection among open field and closed loops of various heights in the CHs. Thus, if the energy dumping events were to occur below the formation height of {\mg}, the upflowing plasma may be accelerated preferentially in CHs, and reach {\si} heights, where a strong correlation is obtained. Similarly, if the event were to occur at much greater heights, or if plasma launched from the lower heights (e.g. type~\rm{II} spicules-like events) are returning to the low solar atmosphere, the downflowing plasma may be falling from {\si} formation height, that will show deceleration due to increasing density towards lower atmosphere mapped by {\mg}. Assuming a mass flux conserving flow, we have $\rho \mathrm{V} = \mathrm{const}$, where $\rho$ is the mass density and $\mathrm{v}$ is the velocity. For a given downflow in {\si}, the downflows in chromospheric lines are smaller in QS over CH (Fig.~\ref{fig:SiMgC_same}.\textbf{a}-\textbf{c}). Hence, the density increase from {\si} formation heights to {\mg} formation heights is larger in QS over CH, resulting in a larger velocity reduction in QS. However, note that since the mass flux is typically different for CHs and QS, a quantitative comparison is beyond the scope of this work. 

For bidirectional flows due to reconnection event between {\mg} and {\si} formation heights, the counterpart upflows will be preferentially accelerated into {\si} formation height in CHs over QS due to excess open expanding flux tubes in CHs over QS. Since the QS has predominantly closed loops, the closed loop reconnection only serves to fill the loop with plasma, raising its intensity. Thus, impulsive events occurring across a range of heights combined with loop statistics in CHs and QS, elegantly ties in together all our observations, and provides a unified scenario for QS heating and solar wind emergence.

Finally, the scenario we present in Fig.~\ref{fig:SiC} is also appealing to explain the switchbacks observed in the near-Sun solar wind \citep{balogh_1999_switchback,Bale_2019_switchbacke} using Parker Solar Probe. One of the competing scenarios for the formation of these switchbacks is through interchange reconnection events occurring in the TR and lower corona \vishal{\citep[][see also \cite{Liang_2021_Zankobs} for an assessment of viability of switchbacks from the linear theory of \cite{Zank_2020_theory}]{fisk_2020_switchbacks,Mozer_2021_switchbackTR, Zank_2020_theory,TriNS_2021,Fargette_2021_switchbacksupergranules,Bale_2021_supergran,Alphonse_2020_minifilament_PSP,Alphonse_2020_jet_psp}}. The kinked-field lines as a result of reconnection between the close loop and open field in the coronal holes, as shown by black arrow in Fig.~\ref{fig:SiC}, may be transported outwards into the solar wind, which are then observed as rotations in the magnetic field. In such a scenario, the flows reported in this paper serve as constraints and modeling inputs for solar wind switchback simulations.

\vishal{A straightforward association between the scenario we present, and polar coronal hole jets is clearly seen. Interchange reconnection seems to play the predominant role in generation of mass flux and plasma heating in these events. However, note that the jets observed by \cite{Moore_2015_untwist_CH} have velocity almost two orders of magnitude more than the velocities we report, and show morphological differences arising due to twist and shear in the ambient magnetic field~\citep[see also][]{Moore_2013_jetrotation}. Since we are averaging over multiple pixels for boosting signal, checking for such morphological signatures is beyond the scope of this work. However, newly-emerged bipoles may interact with the ambient vertical field similar to the scenario proposed by \cite{Moore_2011_xrayjets}, giving rise to Spicule-like events. The interaction height, and amount of magnetic flux converted into thermal energy would then determine the lines which show correlated flow.  }

The observational results and the scenario presented in this paper explain the solar wind formation including switchbacks and the dynamics observed in the QS. We, however, stress that disentangling the absolute effects of wave propagation v/s impulsive upflows is needed. Furthermore, disentangling the effect of return of spicule-like events v/s downflows due to impulsive events occurring higher up in the TR is also difficult. Disentangling these different effects requires further high resolution spectroscopic observations simultaneously taken at different heights, combined with numerical simulations incorporating radiative transfer \& evolution of solar corona into the solar wind. Such observations may be provided with the EUV High-Throughput Spectroscopic Telescope (EUVST) on the upcoming Solar-C mission \citep{Shimizu_2020_solarC_euvst}.

The authors thank T\'eo Bloch (University of Reading) for suggesting binning in deciles and Vilangot Nhalil Nived (Armagh Observatory and Planetarium, U.K ) for providing coalignment and {\si} fitting codes in IDL for cross check. The authors also thank Mats Carlsson (University of Oslo), Mark Cheung (Lockheed Martin Solar and Astrophysics laboratory), Giulio Del Zanna (University of Cambridge), Jim Klimchuk (NASA Goddard)  and Aveek Sarkar (PRL, Ahmedabad, India) for discussions and suggestions. The authors sincerely thank Nishant Singh (IUCAA, Pune, India) for comments on the manuscript. \vishal{U.V. thanks his sister Vaishnavi Upendran for suggestions regarding aesthetics of Fig.~\ref{fig:flow_relations} and Fig.~\ref{fig:SiC}.} We acknowledge the use of data from IRIS, AIA and HMI. IRIS is a NASA small explorer mission developed and operated by LMSAL with mission operations executed at NASA Ames Research Center and major contributions to downlink communications funded by ESA and the Norwegian Space Centre. AIA and HMI are instruments on board SDO, a mission for NASA's Living With a Star program.

\software{Numpy~\citep{numpy_nature},  Astropy~\citep{astropy1}, Sunpy~\citep{sunpy}, Scipy~\citep{scipy},  Scikit-image~\citep{scikit-image}, Matplotlib~\citep{matplotlib}, Multiprocessing~\citep{multiprocessing}, OpenCV~\citep{opencv_library}. Jupyter~\citep{jupyter}.}

\appendix 

\section{Distribution of various properties as a function of {\bmag}}
\label{sec:appdistr}
\begin{figure*}[htpb!]
  \centering
  \includegraphics[width=\textwidth]{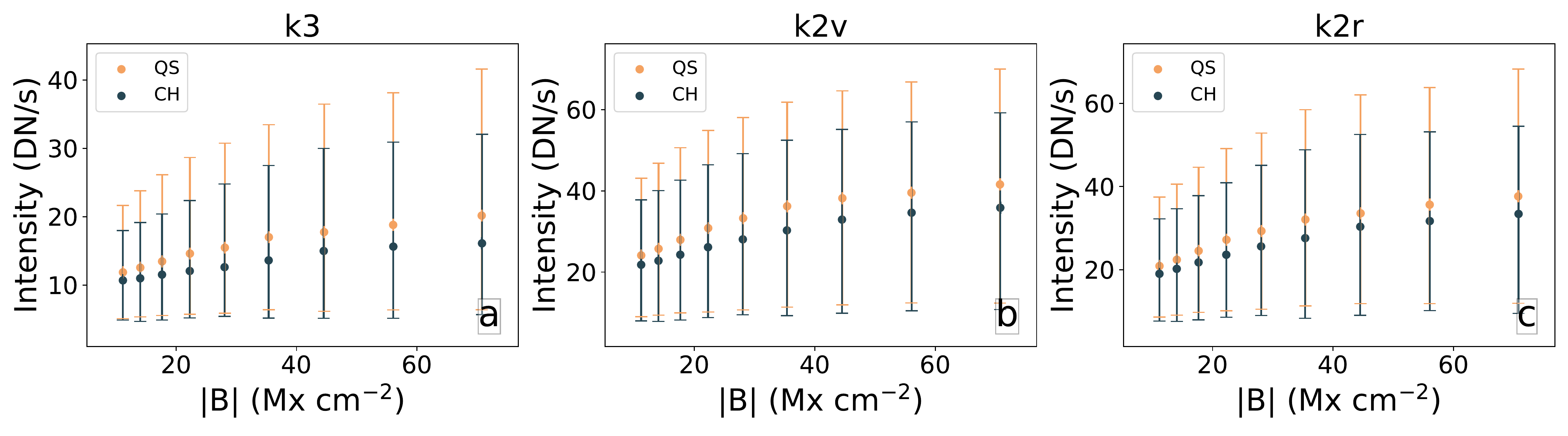}
  \caption{Same as Fig.~\ref{fig:mg2_intens_comb}, but the errors now represent $1$ and $90$ percentile bounds of the distribution of samples present in the bin of {\bmag}.}
  \label{fig:mg2intensdistr}
\end{figure*}

\begin{figure*}[htpb!]
  \centering
  \includegraphics[width=\textwidth]{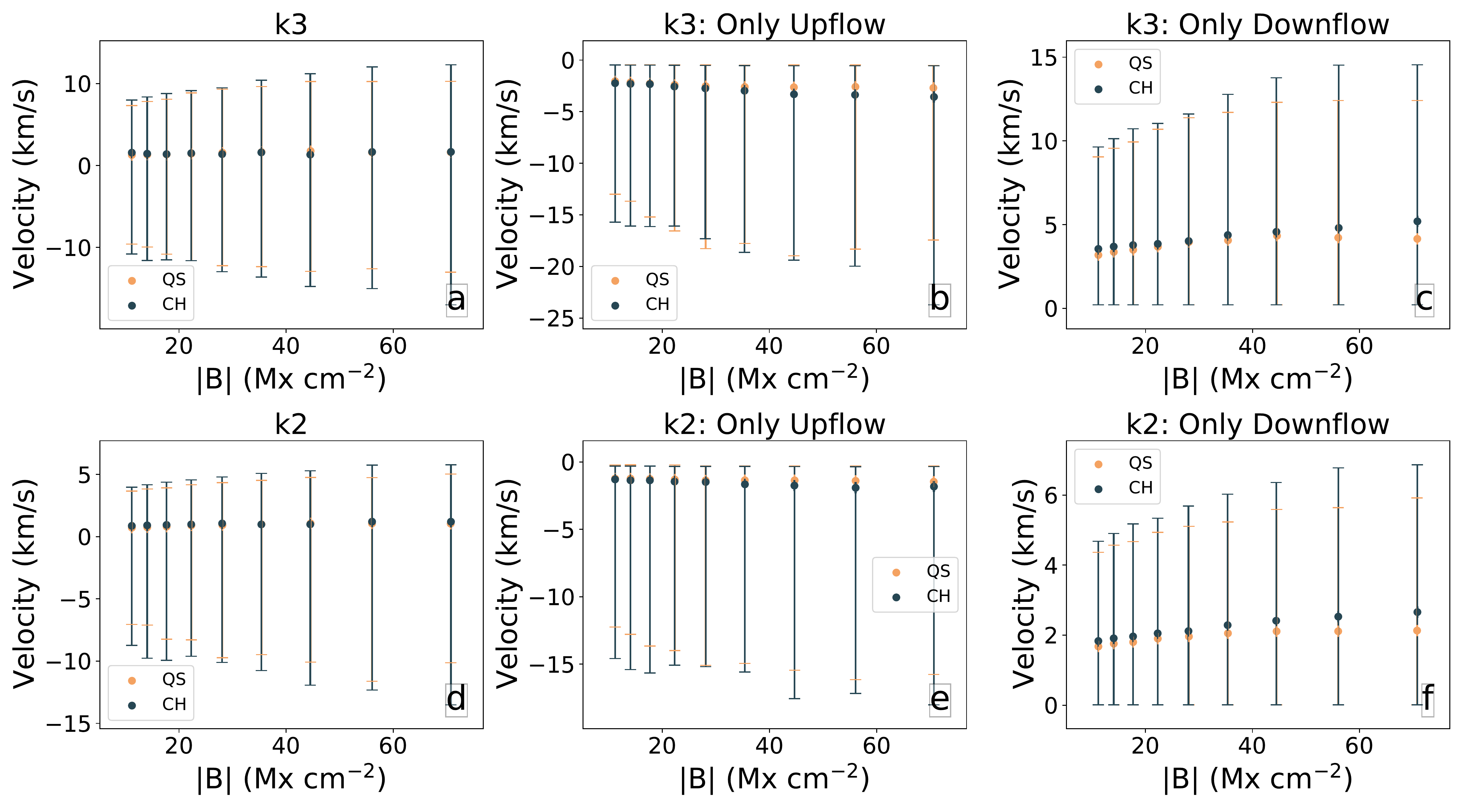}
  \caption{Same as Fig.~\ref{fig:mg2_velocity_comb}, but the errors now represent $1$ and $90$ percentile bounds of the distribution of samples present in the bin of {\bmag}.}
  \label{fig:mg2veldistr}
\end{figure*}

\begin{figure*}[htpb!]
  \centering
  \includegraphics[width=\textwidth]{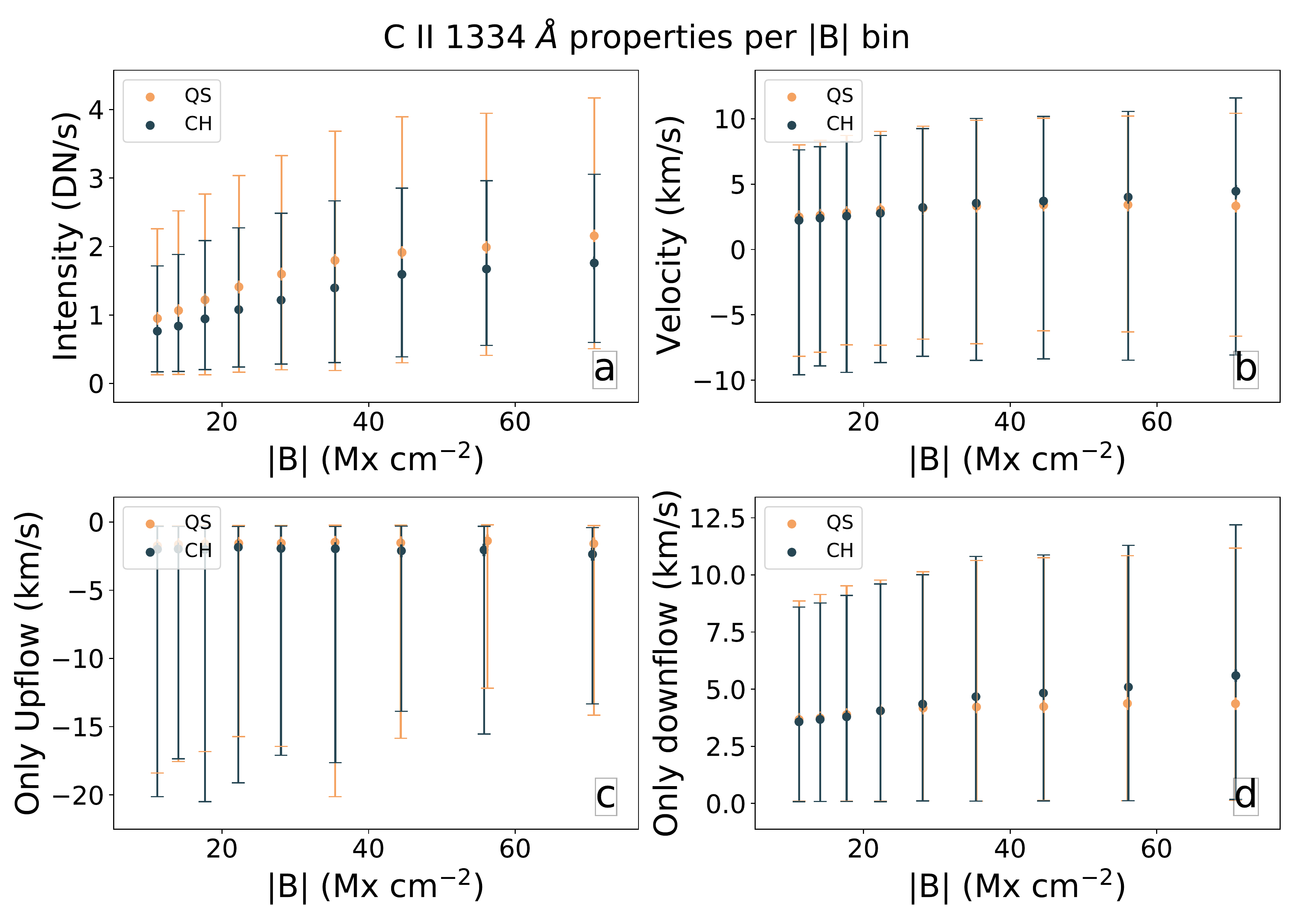}
  \caption{Same as Fig.~\ref{fig:c2prop}, but the errors now represent $1$ and $90$ percentile bounds of the distribution of samples present in the bin of {\bmag}.}
  \label{fig:c2distr}
\end{figure*}

\begin{figure*}[htpb!]
  \centering
  \includegraphics[width=\textwidth]{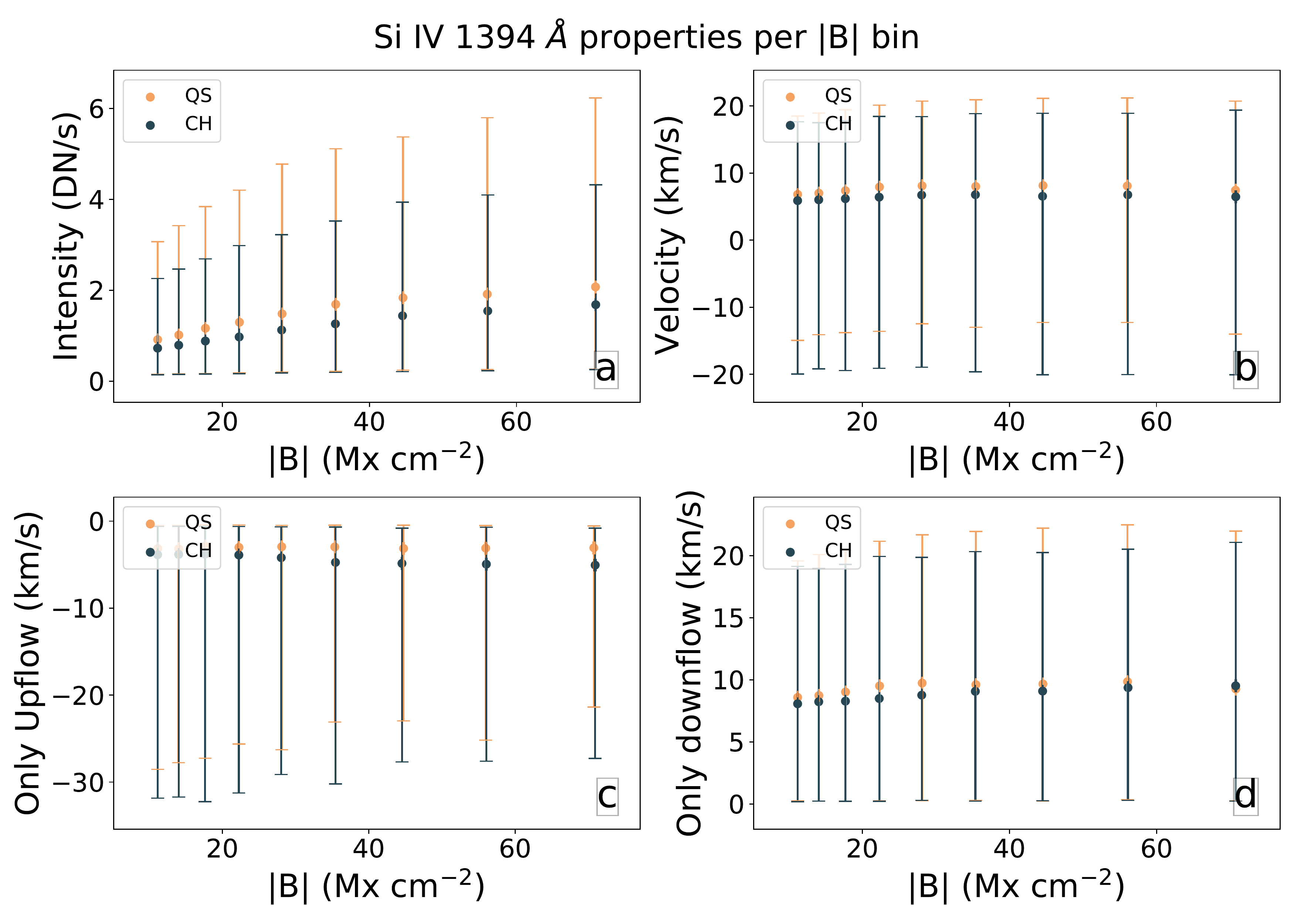}
  \caption{Same as Fig.~\ref{fig:si4prop}, but the errors now represent $1$ and $90$ percentile bounds of the distribution of samples present in the bin of {\bmag}.}
  \label{fig:si4distr}
\end{figure*}

We present here the distribution of intensity and velocity from {\mg}, {\car} and {\si} lines as a function of {\bmag}. The plots shown from Fig.~\ref{fig:mg2intensdistr} - \ref{fig:si4distr} are the same as those from Fig.~\ref{fig:mg2_intens_comb} - \ref{fig:si4prop}, except that the errorbars reported correspond to $1$ and $90$ percentile of the samples in each bin.

Since the overall distribution of the various quantities look very similar in CHs and QS, the distribution within bins of {\bmag} reflects how systematically differences arise in the ensemble of pixels considered. The distribution of samples between CH and QS slowly drift apart in their mean values, depicting the transition from statistical signatures in the lower atmosphere to very clear signatures in the corona. Thus, the distribution of samples in each bin provides further constrains to expected results from simulations.


\bibliography{manuscript}{}
\bibliographystyle{aasjournal}

\end{document}